\documentclass[aps,prb,showpacs,superscriptaddress,amsmath,twocolumn]{revtex4-1}

\usepackage{amsmath}
\usepackage{amssymb}
\usepackage{graphicx}
\usepackage[english]{babel}
\usepackage{natbib}

\usepackage{epstopdf}

\begin{document}

\title{Quantum interference and nonequilibrium Josephson current in molecular Andreev interferometers}

\author{Noel~L.~Plaszk\'o}
\affiliation{Dept. of Physics of Complex Systems, E\"otv\"os Lor\'and University, Budapest, P\'azm\'any P. s. 1/A, Hungary}

\author{Peter Rakyta}
\affiliation{Dept. of Physics of Complex Systems, E\"otv\"os Lor\'and University, Budapest, P\'azm\'any P. s. 1/A, Hungary}

\author{J\'ozsef Cserti}
\affiliation{Dept. of Physics of Complex Systems, E\"otv\"os Lor\'and University, Budapest, P\'azm\'any P. s. 1/A, Hungary}

\author{Andor Korm\'anyos}
\affiliation{Dept. of Physics of Complex Systems, E\"otv\"os Lor\'and University, Budapest, P\'azm\'any P. s. 1/A, Hungary}

\author{Colin~J.~Lambert}
\affiliation{Dept. of Physics, Lancaster University, Lancaster, LA1 4YB, United Kingdom}

\begin{abstract}
We study the  quantum interference (QI) effects in three-terminal Andreev interferometers based on polyaromatic hydrocarbons  (PAH's) under non-equilibrium conditions. The Andreev interferometer consists of a PAH coupled to two superconducting and one normal conducting terminals. 
We calculate the current measured in the normal lead as well as the current between the superconducting terminals under non-equilibrium conditions. 
We show that both the  QI arising in the PAH cores and the bias voltage applied to a normal contact have a fundamental effect on the charge 
distribution associated with the Andreev Bound States (ABS's).  QI  can lead to a peculiar dependence of the normal current on the superconducting 
phase difference that was not observed in earlier studies of  mesoscopic Andreev interferometers. 
We explain our results by an induced asymmetry in the spatial distribution of the electron- and hole-like quasiparticles. 
The non-equilibrium charge occupation induced in the central PAH core can result in  a  $\pi$ transition in the  current-phase relation 
of the supercurrent for large enough applied bias voltage on the normal lead.
The  asymmetry in the spatial distribution of the electron- and hole-like quasiparticles might be used 
to split Cooper pairs and hence to produce entangled electrons in four terminal setups.
\end{abstract}

\maketitle

\section*{Introduction} \label{sec:introduction}

{Quantum interference (QI) is ubiquitous in nature. Constructive quantum interference (CQI) leads to the formation of energy levels 
in atoms or molecules and energy bands in crystals, whereas destructive quantum interference (DQI) leads to energy gaps in molecules and band gaps in solids. 
The energy scale for these QI phenomena can be up to a few $eV$ and therefore these quantum effects control the properties of 
molecules and solids at room temperature, for which $k_B T\approx 25 \mbox{meV}\ll 1 \mbox{eV}$. 
In addition to these high temperature manifestations of QI, many low-temperature interference phenomena are well known, such as superfluidity and 
superconductivity, which occur on energy scale of order a few $meV$ or less. }

{Investigations of QI in condensed systems are often driven by the desire to harness QI and deliver useful function. 
For example, when a molecule is placed into the nanogap between two metallic electrodes, it is known that electron transport 
from the source to the drain electrode is phase coherent at room temperature, provided the length of the molecule is less than approximately $3$nm. 
Consequently, if the interference pattern created by electronic de Broglie waves passing through the molecule can be controlled, 
then useful room-temperature devices such as molecular-scale switches, transistors and sensors could be realised. 
Single-molecule electronics is the sub-field of nanoelectronics\cite{Cuevas-Sheer,Ratner2013,Tsutsui2012,Loertscher2013,Poulsen2014,Nuckolls2016,GuoXuefeng2016}, 
which aims to deliver such structures and in pursuing this goal,
many groups have demonstrated that electrons can be injected into (and collected from) the core of a molecule with atomic accuracy\cite{Venkataraman2016,Gehring2019,HongWenjing2019,HuWenping2019}. 
Furthermore, it has been demonstrated that an ability to vary the atomic-scale connectivity to molecular cores is an effective way of 
controlling room-temperature QI\cite{Sangtarash2015,Sangtarash2016}. On the other hand, at lower temperatures, quantum engineers strive to utilise QI in superconducting 
structures such as SQUIDs and Andreev interferometers, which rely on controlling the interplay between a superconducting condensate 
and charge-carrying quasi-particles\cite{Lambert1993,Hui1993,PhysRevLett.95.147001,Giazotto-APL,vanWees-theory,Bagwell-threeterminal,Baselmans-pijunction,PhysRevLett.89.207002,PhysRevB.77.014528}. In such devices, QI is controlled by the phase of the superconducting order parameter, 
which describes a macroscopic collective degree of freedom, which has no counterpart at room temperature.}

{In this article, we examine the interplay between the high-energy-scale QI found in molecules and the low-temperature QI present in superconductors. 
Our aim is to determine how an ability to control the connectivity to molecular cores with atomic accuracy can be used to engineer 
the properties of Andreev interferometers and SQUIDS. }

{From the viewpoint of connectivity, a fundamental manifestation of QI is illustrated in Fig.~\ref{fig:setup} top and middle panels, which shows an 
anthanthrene molecular core (consisting of $6$ six-membered rings) connected by triple bonds to external electrodes. 
The connectivity of the triple bonds to the core is fixed by chemical synthesis. Fig.~\ref{fig:setup} shows two examples of molecules with different connectivities. 
Following the numbering scheme of the lattice shown at the bottom of Fig.~\ref{fig:setup}, molecule $1$ has triple bonds connected to atoms 
$12$ and $3$, whereas molecule $2$ has triple bonds connected to atoms $9$ and $22$. The triple bonds are connected to terminal aryl rings, 
which in turn are connected to thioacetate anchor groups, which bind the molecules to source and drain electrodes. 
Since the triple bonds form weak links to the central core, it is conceptually convenient to consider the combination of an aryl ring, anchor group 
and external electrode as a single ``compound electrode'', (coloured blue in Fig.~\ref{fig:setup}) which injects or collects electrons 
to or from the central core, via the triple bonds. Remarkably, when the external electrodes are normal (i.e., not superconducting), 
the room-temperature electrical conductance of setup $1$ is both measured and predicted to be almost two orders of magnitude higher than that of $2$. 
This is a clear manifestation of room-temperature QI, since the conductance of a corresponding classical resistor network formed from six rings 
of resistors would show a much lower dependence on connectivity. From the viewpoint of superconductivity, our aim is to replace one or 
more of the normal electrodes by superconducting electrodes and examine how electron transport though such molecular cores is controlled 
by a combination of connectivity and by the phase of the superconducting order parameter.}

In ballistic normal-superconductor (N-S) hybrid systems the fundamental transport process is  Andreev reflection,
whereby an incoming electron is reflected back as a hole at the N-S interface. 
A rich set of physical phenomena that follow from this scattering process  was realized in Andreev interferometers, 
which are devices with two (or more) superconducting and one (or more) normal leads attached to a central region\cite{Lambert1993,Hui1993}.
For example, due to the extraordinary sensitivity of the Andreev current to the superconducting phase difference,  Andreev interferometers may provide a 
faster and more precise alternative to  superconductor quantum interference devices  (SQUIDs)\cite{PhysRevLett.95.147001} to 
measure properties of quantum systems or even detecting Majorana bound states\cite{Giazotto-APL}. 
Importantly, the presence of a normal lead allows one to change the equilibrium occupation of Andreev bound states formed in multi-terminal
N-S systems. It was suggested that such a non-equilibrium effect  can be used to engineer $\pi$-Josephson junctions\cite{vanWees-theory, Bagwell-threeterminal}, 
where the fundamental relation  $I_s=I_c\sin(\delta\phi)$ between the phase difference 
$\delta\phi$ of the order parameters of two superconductors and the supercurrent $I_s$ can be changed  to $I_s=I_c\sin(\delta\phi+\pi)$ 
($I_c$ is the critical current).  This effect has indeed been measured in diffusive meso-scopic multi-terminal N-S 
systems\cite{Baselmans-pijunction,PhysRevLett.89.207002,PhysRevB.77.014528}.

Recently, the  superconducting properties of molecular-scale 
junctions have also started to attract experimental\cite{C60-supercond,nature-andreev-qubit,Weber2017,PhysRevLett.118.117001} as well as 
theoretical\cite{PhysRevB.79.075119,Nappi2018,Rakyta2019} interest. In  Ref.~\cite{Rakyta2019} we  discussed the equilibrium properties of 
various multiterminal N-S systems where, in particular, QI effects in the core of the molecule play an important role.  
Here we show how such QI effects and  non-equilibrium charge injection can lead to interesting effects in molecular Andreev interferometers.  
Namely, the non-equilibrium occupation of the Andreev bound states (ABS's) which are formed in superconductor-molecule-superconductor (S-M-S')
Josephson junctions can be driven via the third, normal lead attached to the Josephson junction, thus realizing a non-equilibrium N-M-SS' system. 
As already mentioned, one of the key ingredients in our work is QI  which arises in the molecular core of  N-M-SS' systems that  are based on 
PAHs\cite{Geng2015,Sangtarash2015,Sangtarash2016}. We find that in these systems one may observe effects that were
not attainable is previously studied mesoscopic Andreev interferometers.   
Based on the ``magic number theory of  connectivity in Refs.~\cite{Geng2015,Sangtarash2015,Rakyta2019} we  show that conductive channels through 
the molecular core can give rise to interfering paths contributing to the total ABS wave function with the same or with an opposite 
sign for electron and hole-like degrees of freedom. 
This rich set of interfering paths is provided by the conductive channels opened by the insertion of a substituent heteroatom into the molecular core\cite{Sangtarash2016}.
Under specific circumstances the interplay of the interfering amplitudes may even lead to the total suppression of the electron-like (or hole-like) 
degrees of freedom on certain molecular sites and, at the same time, to a constructive interference for the hole-like (or electron-like) charge carriers.
Since the charge current through the normal lead is closely tied to the Andreev reflection process, its magnitude is highly influenced by 
the density of both the electron- and hole-like particles in the vicinity of the normal contact. 
Thus, by measuring the charge current through the normal lead one can also probe the electron-hole separation in the molecular junction.

In what follows, we first describe the main characteristics of  interference effects in Andreev interferometers
based on  PAHs\cite{Geng2015,Sangtarash2015,Sangtarash2016} in equilibrium conditions.  
Our choice is justified by the peculiar mid-gap transport properties of these molecules accompanied by inner quantum interference effects within the core of the molecule\cite{Sangtarash2015,LambertChemSocRev,Sangtarash2016,Sadeghi2658,Sedghi2011,Zhao2013,PhysRevB.74.193306,doi:10.1063/1.3451265,Vazquez2012,PhysRevLett.109.056801,Aradhya2012,doi:10.1002/anie.201207667,Guedon2012,Manrique2015,Rakyta2019}.
We outline an illustrative connectivity-based theory that can be used to understand the current-phase relations at non-equilibrium conditions as well.
Then we present our numerical results obtained for the normal and for the supercurrent at finite bias applied on the normal lead. 
We interpret our results in terms of connectivity arguments. 
We examine how the electrical properties of the Andreev interferometers would be influenced by tuning the inner QI effects of the molecular core.
We demonstrate how QI can lead to a suppression of the normal current which is a clear evidence of the spatial separation of the electron- and hole-like particles.
Finally,  we present a summary of our most important results and give a brief outlook.

\section*{Theoretical background: equilibrium molecular Andreev interferometers}
\label{sec:equilibrium}

\begin{figure}[h]
\begin{centering}
\includegraphics[trim={0 0 0 0},clip, width=0.95\linewidth]{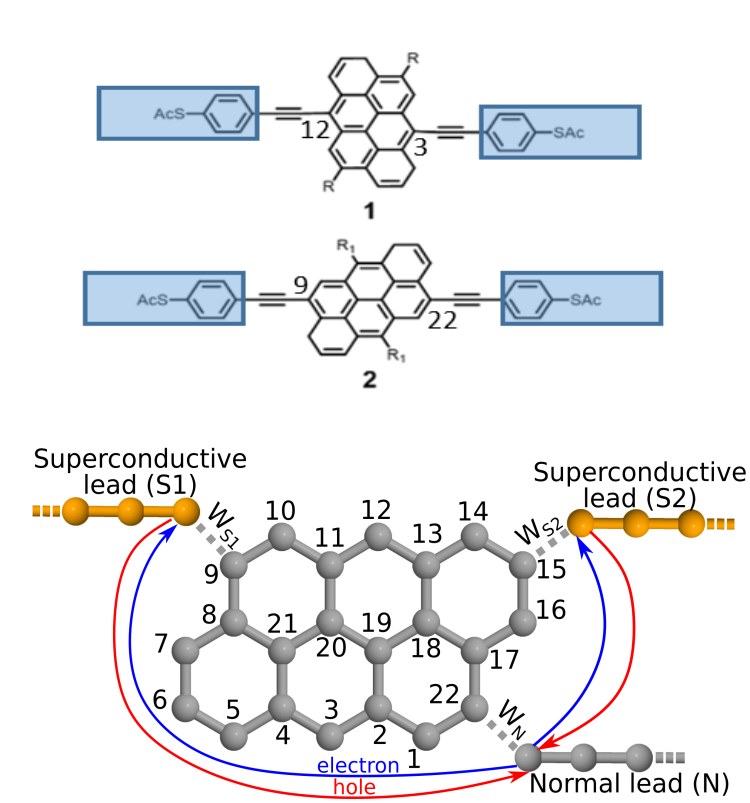}
\caption{The top and middle panels show molecules $1$ and $2$, with connectivities $12,3$ and $9,22$ to the anthanthrene molecular core.
The lower panel shows an Andreev interferometer consisting of an anthanthrene molecule, two superconducting leads and one normal lead. 
The ``sites'' of the associated tight-binding model represent p$_z$  orbitals of the anthanthrene molecule and are  labelled according to the figure. 
The coupling  of the molecule to the  normal (superconducting) lead is denoted by $W_N$ ($W_S$), for details see the text. 
The superconducting leads are characterized by a superconducting order parameter $\Delta e^{i\phi_1}$ and $\Delta e^{i\phi_2}$. 
The transport processes responsible for the conventional interference effect are indicated by solid 
lines for the electron-like (blue) and hole-like (red) propagation.}
\label{fig:setup}
\end{centering}
\end{figure}

From a conceptual viewpoint the key ingredients of our theoretical model are based on  weak coupling, connectivity-driven, 
mid-gap transport and phase coherence. A detailed explanation  of these assumptions is given in Refs.\cite{Geng2015,Sangtarash2015,Rakyta2019}.
Here we only mention that  the term ``weak coupling'' means that the central aromatic molecule is weakly coupled to the contacts 
resulting in a small level broadening and self energy correction to the HOMO and LUMO levels compared to the HOMO-LUMO gap.
Thus, provided the Fermi level lies within the gap (resulting in near mid-gap transport), the quantum interference effects in the 
phase coherent transport processes are characterized by the properties of the molecular core alone.
(The Coulomb interactions can be included at the level of a self-consistent mean field description such as
Hartree, or Hartree-Fock.)
Taken together, these conditions ensure that when computing the Green's function of the
core, the contribution of the electrodes can be ignored.
Consequently, the probability of the propagation of the charged particles between sites $k$ and $l$ of the molecule is described by the 
``magic number'' matrix element\cite{Geng2015,Sangtarash2015,Rakyta2019} $M_{kl}\sim g_{kl}$, 
where $g(E_F) =\left( E_F-H_{\textrm{mol}} \right)^{-1}$ is the Green's function of the isolated molecule described by Hamiltonian $H_{\textrm{mol}}$. 
In particular, the electrical conductance corresponding to connectivity $k,l$ is proportional to $|M_{kl}|^2$.
In the simplest theoretical description an  integer valued connectivity matrix  $H_{\textrm{mol}}$ captures the complexity of the inner CQI and DQI
 effects within the core of the molecule and when $E_F$ coincides with the middle of the HOMO-LUMO gap, the 
resulting magic number matrix $M_{kl}$ is simply a table of integers. 
In particular, for the bipartite lattice of Fig.~\ref{fig:setup}, when  $E_F$ coincides with the middle of the HOMO-LUMO gap, 
destructive quantum interference arises between sites  $k$ and $l$ that have the same parity (i.e., both are odd or both are even)  
and therefore the matrix element $M_{kl}$ is zero. In contrast, when the  sites  $k$ and $l$  have different parity,  $M_{kl}$ may be finite giving a
non-zero propagation amplitude of the charged particles between  sites  $k$ and $l$. For the anthanthrene core of Fig.~\ref{fig:setup}, 
$M_{3,12}=1$ and $M_{9,22}=9$. Hence their conductance ratio is predicted to be $|M_{9,22}|^2/|M_{3,12}|^2=81$, which is close to the measured value
of the conductance ratio, both for single molecules and for self-assembled monolayers\cite{Geng2015,Sangtarash2015,Famili2019}.

In order to understand the unconventional non-equilibrium Andreev interference effect described in the next sections, 
following  Ref.\cite{Rakyta2019} we first give a brief overview of the considerations that explain the interference pattern 
in the  current $I_N$ flowing through the normal lead  as a function of the superconducting phase difference $\delta\phi=\phi_1-\phi_2$ 
between the S1 and S2 leads (see Fig.~\ref{fig:setup}) in equilibrium. 
Under equilibrium conditions, the limit $eV\rightarrow 0$ is understood, where $V$ is the applied bias on lead N 
with respect to the chemical potential of the superconductors. 
Note that the imposition of a phase difference $\delta\phi$  also generates a Josephson current  $I_s$ 
flowing between the superconducting leads. 
Experimentally, as shown in Refs\cite{UrbinaC2007,Nanda2017},  $\delta\phi$  can be controlled, thus allowing the measurement
of the current-phase relation (CPR) of the supercurrent $I_s$ and the phase dependence of $I_N$.

Let us investigate $I_N$ in a device consisting of an anthanthrene central molecular core, as shown in Fig. \ref{fig:setup}. 
The  $\delta\phi$ dependence of $I_N$ can be understood as a result of an interference effect between the possible transport paths 
of electrons and holes.  
 The arms of the interferometer are formed by the 
 trajectories $\textrm{N}\rightarrow \textrm{mol}\rightarrow \textrm{S1}\rightarrow \textrm{mol} \rightarrow \textrm{N}$ and 
 $\textrm{N}\rightarrow \textrm{mol}\rightarrow \textrm{S2}\rightarrow \textrm{mol} \rightarrow \textrm{N}$.
 Let us  consider the setup in which the normal lead $N$ is attached to site labeled by $22$, and the superconducting leads $S1$ and $S2$ are attached to sites $9$ and $15$, respectively and examine the transmission amplitude $t_{9,22}$ related to the process $\textrm{N}\rightarrow \textrm{mol}\rightarrow \textrm{S1}\rightarrow \textrm{mol} \rightarrow \textrm{N}$.
Since the normal reflection on the superconductors does not give a contribution to the charge current, only Andreev reflection\cite{Andreev1964}
can cause a net charge current. During Andreev reflection, an incoming electron-like quasiparticle is converted into a hole-like quasiparticle and vice versa 
at the normal-superconductor interface. 
Due to the Andreev reflection, an extra $e^{-{\rm i}\phi_1}$ phase factor multiplies the transmission amplitude 
($\phi_1$ is the phase of the superconductor S1). 
Then, the reflected hole-like state propagates back to the normal lead, a 
process which can be described by $-M_{9,22}$ according to the Bogolioubov-de Gennes equation\cite{Rakyta2019}.
Based on these considerations, the transmission amplitude can be written in the following form:
\begin{equation}
    t_{9,22} \sim -M_{9,22}^2 e^{-{\rm i}\phi_1}. \label{eq:tkl}
\end{equation}
Similar considerations can made for the transmission amplitude $t_{15,22}$. 
Since there are two interfering arms in the interferometer, one needs to sum up both transmission amplitudes associated with 
the two propagation paths to calculate the total transmission amplitude:
\begin{equation}
 t_{tot} \sim t_{9,22} + t_{15,22} = -M_{9,22}^2 e^{-{\rm i}\phi_1} -M_{15,22}^2 e^{-{\rm i}\phi_2} ,
\end{equation}
From this expression the Andreev current $I_N$ at small bias voltage ($eV\ll\Delta$) can be approximated as:
\begin{equation}
 I_N \sim |t_{tot}|^2 = M_{9,22}^4 + M_{15,22}^4 + 2 \cdot M_{9,22}^2 \cdot M_{15,22}^2 \cdot \cos (\phi_1-\phi_2).
 \label{eq:In-equilibrium}
\end{equation}
As one can see, the Andreev current $I_N$ is indeed expected to show a simple dependence on the superconducting phase difference 
$\delta\phi=\phi_1-\phi_2$ with a minimum at $\pi$.
Regarding the supercurrent $I_s$ flowing between S1 and S2, in first approximation, this can be understood as a consequence 
of Andreev bound states (ABS), although a continuum of unbound states can also add a finite contribution\cite{PhysRevB.46.12573}.


\section*{Non-equilibrium numerical calculations}
\label{sec:nonequibl-methods}

To avoid time-dependent order parameter phases varying at the Josephson frequency, we assume  that the superconductors 
S1 and S2 share a common condensate chemical potential $\mu$. A finite bias voltage $V$ (with respect to $\mu$) can be then 
applied to the normal lead.  This bias voltage will affect both the normal current $I_N$  and, by changing the equilibrium occupation of the ABSs,
the supercurrent $I_s$ as well.  

{In order to describe the transport properties at finite bias voltage  one has to use a theoretical 
framework capable of describing non-equilibrium transport processes.
We calculate the  currents  $I_N$ and $I_s=(I_{S_1} - I_{S_2})/2$ by using a tight binding approach and the
Keldysh non-equilibrium Green's function techniques\cite{PhysRevB.68.075306,NEGFtheory,Pala.2007}:
\begin{equation}
 I_{N(S_i)} = -\frac{2e}{h}{\rm Re}\left[\int dE \;{\rm Tr} \left( \tau_3\Gamma_{N(S_i)} G^{<}(E)\right )\right], \label{eq:current}
\end{equation}
with $\Gamma_{N(S_i)}$ being the coupling from the molecule to the normal (superconducting) lead labeled by $N$ ($S_i$) including the electron-hole degrees of freedom and $\tau_3$ is the third Pauli matrix acting on the electron-hole space.
The current $I_{N(S_i)}$ describes the current flowing through lead $N$ ($S_i$) into the central molecule. 
In the steady state limit the currents flowing through the individual leads satisfy the charge conservation rule $I_{N} + I_{S_1} + I_{S_2} = 0$ leading to two independent currents $I_N$ and $I_s$ characterizing the electrical properties of the junction.
Finally, the lesser Green's function $G^<$ in Eq.(\ref{eq:current}) can be calculated within the Keldysh non-equilibrium framework using the 
Keldysh equation (see details in the electronic supporting information).
The calculations were performed using the tight-binding framework implemented in the EQuUs\cite{EQUUS} package.
The relevant electronic states in the molecular core were described by a single orbital tight-binding model where the nearest neighbor sites are connected 
by a hopping amplitude $\gamma_0$. 
As shown in Fig.~\ref{fig:setup}, the hopping amplitude between the molecule and the normal $N$ (superconducting $S1$, $S2$) lead 
is given by $W_N$ ($W_{S1}$, $W_{S2}$).  In our calculations, unless indicated otherwise, we used $W_N=0.1\gamma_0$ and $W_S=0.3\gamma_0$.  
The normal and superconducting contacts were modeled by a one-dimensional tight-binding chain. The magnitude of the superconducting order parameter in the 
leads $S1$ and $S2$ was $\Delta=3\cdot 10^{-3}\gamma_0$. The results that we are going to discuss do not depend on the actual value of $\gamma_0$ and $\Delta$. 
This simple model is justified by previous studies of connectivity driven transport processes through PAH molecules\cite{Geng2015,Sangtarash2015,Rakyta2019}.
Following these works, our aim is to highlight the role of connectivity in the transport properties of these molecular cores leading to new interference phenomena.
We give the remaining details of the tight binding-model used in our calculations in the electronic supplementary material.}


\section*{Non-equilibrium molecular Andreev interferometers}
\label{sec:nonequilibrium-results}

\begin{figure}[h]
\begin{centering}
\includegraphics[trim={0 0cm 0 0cm},clip, width=0.75\linewidth]{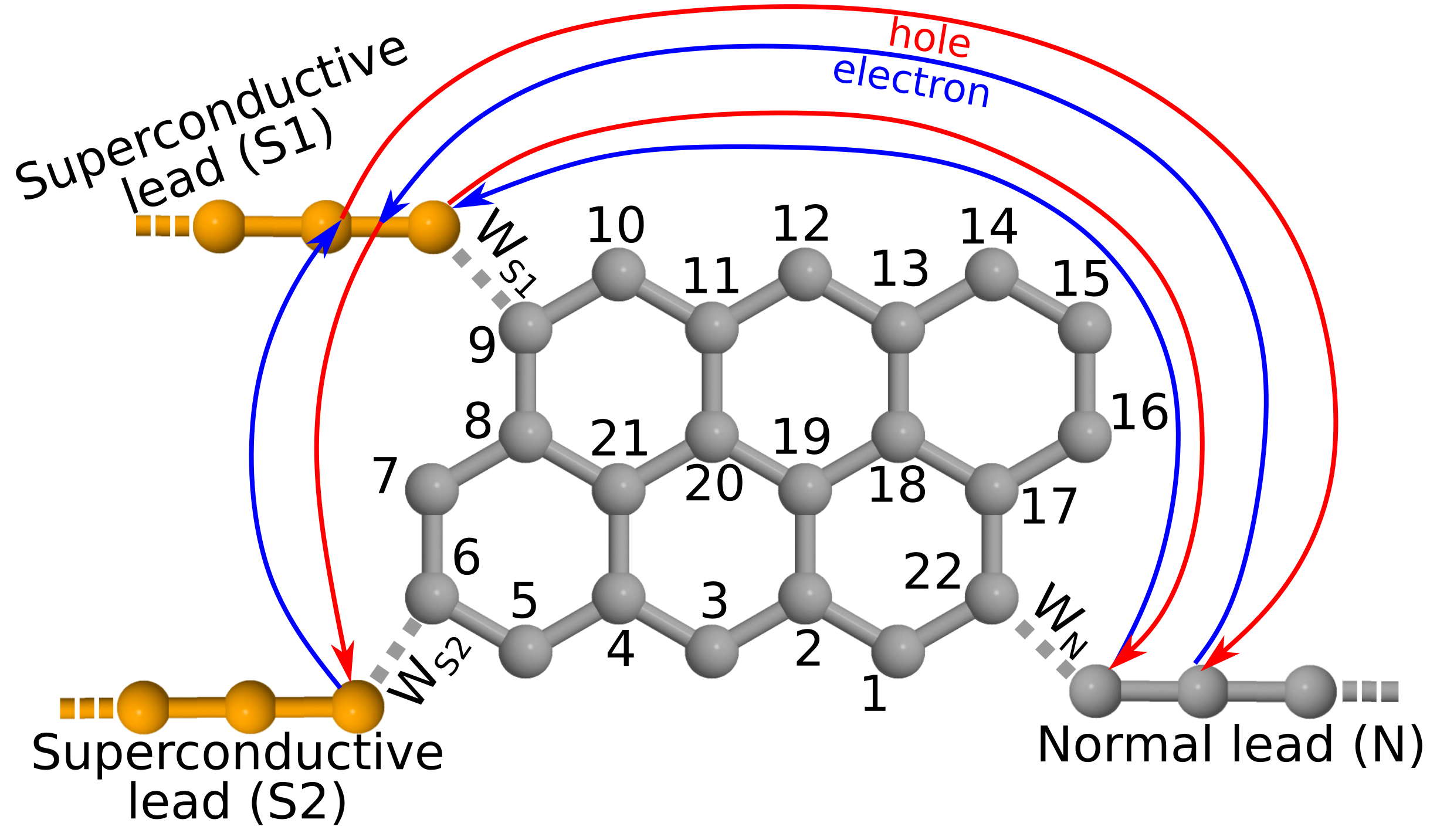}
\caption{Anthanthrene molecule attached to two superconductive and one normal lead. The connectivity matrix element between 
the sites $6$ and $22$ is zero, while the connectivity between sites $9$ and $22$ and between sites $9$ and $6$ is finite.
Solid lines indicate the propagation of  the electron-like (blue) and hole-like (red) quasiparticles.}
\label{fig:ABS-anthanthrene}
\end{centering}
\end{figure}

\begin{figure}[ht]
\begin{centering}
\includegraphics[scale=0.3]{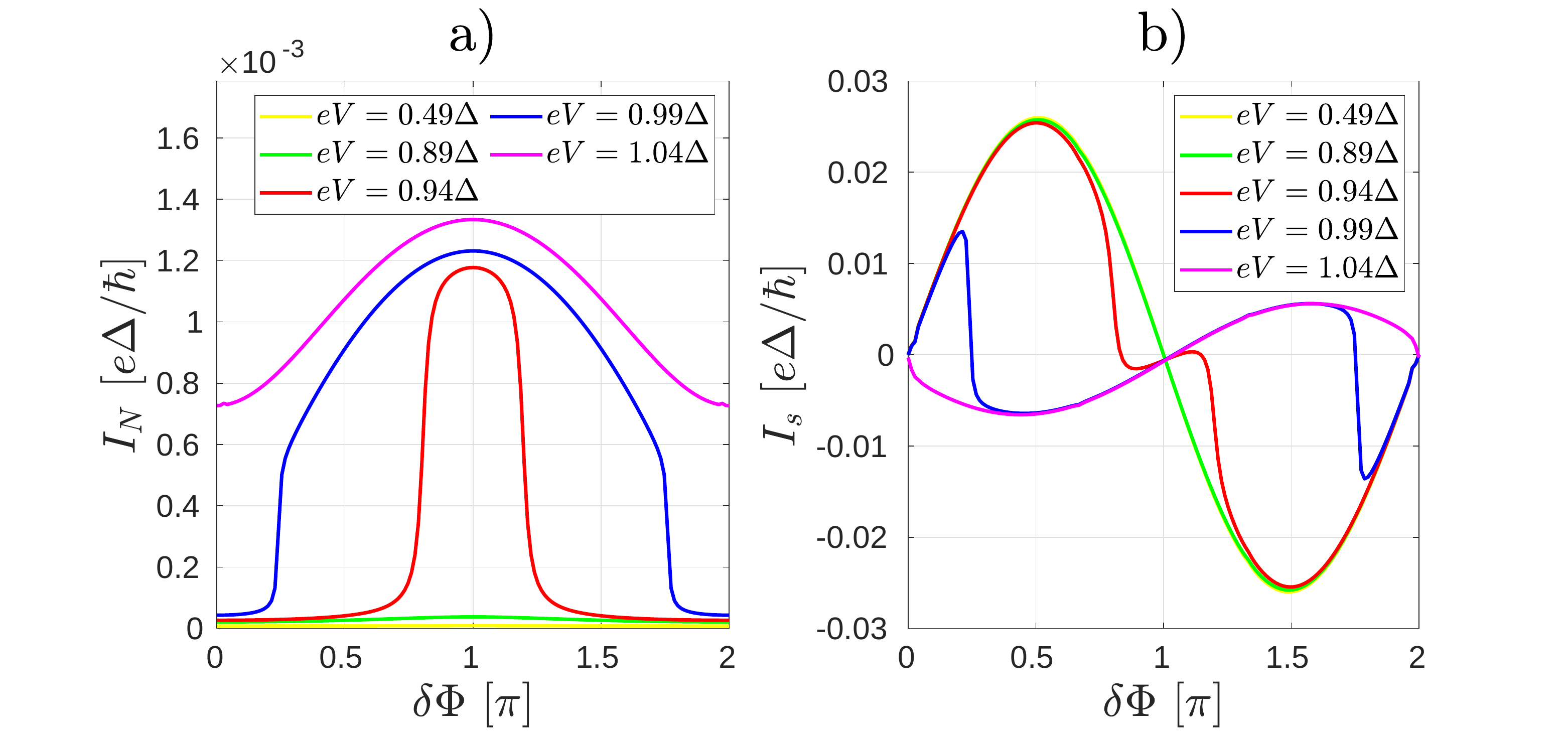}
\caption{The currents $I_N$ (a) and $I_s$ (b) as a function of the phase difference $\delta\phi$ between the superconducting leads 
for the system  depicted in Fig. \ref{fig:ABS-anthanthrene} for several bias voltage $eV$.  
a) a robust peak in $I_N$ appears   around the phase the difference $\pi$  when the bias voltage is comparable 
to the superconducting gap $\Delta$. 
b) the supercurrent shows a $\pi$ transition for $eV>\Delta$. 
}
\label{fig:ABS-anthanthrene-currents}
\end{centering}
\end{figure}

As a first example of non-equilibrium effects in Andreev-interferometers it is instructive to consider the system shown 
in Fig.\ref{fig:ABS-anthanthrene}. With respect to Fig.~\ref{fig:setup},  we   changed  the connecting sites of the leads 
in order to ``disarm'' one of the interfering arms. 
This can be achieved by choosing  connecting sites such that the magic number matrix elements between the sites connected to the normal lead
and to one of the superconducting leads becomes zero, as shown in Fig.~\ref{fig:ABS-anthanthrene}. Therefore 
one may expect   $I_N$   to be independent of the superconducting phase.  Note, however, that  the magic number $M_{6,9}$ between the 
superconducting leads is finite. Therefore, as we will show later,  an ABSs is  formed in this system  and it has an important 
effect on $I_N$.  The results for $\delta\phi$ and $eV$ dependence of $I_N$ can be seen in Fig.\ref{fig:ABS-anthanthrene-currents}(a). 
 The current $I_N$ remains very small for applied voltages  $eV\ll \Delta$ on the normal lead. 
As  $eV$ is increased, a finite $I_N$ starts to flow, but in contrast to the $\sim\cos\delta\phi$ dependence 
given in Eq(\ref{eq:In-equilibrium}),  $I_N$ exhibits a \emph{maximum} at superconducting phase difference $\delta\phi=\pi$. 

These results can be explained by the effect of an ABS.  As pointed out in earlier 
studies on multiterminal normal-superconductor mesoscopic systems\cite{vanWees-theory,Bagwell-threeterminal},  
the voltage $eV$ sets the effective electrochemical potential for the ABSs and those with energy $E_{n,ABS}(\delta\phi)\le eV$ are filled.   
The ABS energy $E_{n,ABS}(\delta\phi)$ depends on the phase difference $\delta\phi$. 
A change  in the occupation of the ABSs  directly affects  $I_{S1}$ and $I_{S2}$ and 
therefore the current distribution in the Andreev-interferometer junction will depend on both the voltage $eV$ and 
on the phase difference $\delta\phi$. 
To illustrate these effects we show  the supercurrent   $I_s=(I_{S1}-I_{S2})/2$  in Fig\ref{fig:ABS-anthanthrene-currents}(b). 
As $eV$ is increased, a deviation from the simple $I_s=I_c\sin(\delta\phi)$ relation can be clearly seen and  
 for $eV>\Delta$  a $\pi$-transition takes place in $I_s$, similarly to what was obtained in Refs.~\cite{vanWees-theory, Bagwell-threeterminal}.

One can  give a heuristic argument of  why  the presence of ABSs can affect $I_N$. This argument draws on analogies  with the 
discussion given for the equilibrium case, i.e., it is based on interfering quasiparticle trajectories.
Although in the system depicted in Fig.\ref{fig:ABS-anthanthrene} the connectivity between the normal
lead and the superconducting lead $S2$ is zero,  the charge carriers can still  probe
 the phase of lead $S2$ when they propagate along a path that also includes an Andreev reflection from the lead $S1$. 
 Namely,  as illustrated in Fig.~\ref{fig:ABS-anthanthrene}, both $M_{22,9}$ and $M_{6,9}$ are finite. 
 We denote the amplitude of such propagation by  $t^{(9)}_{6,22}$, where the upper index $(9)$ indicates that the propagation 
 between the sites $6$ and $22$ takes place via  site the $9$.  
 To approximate the  amplitude $t^{(9)}_{6,22}$ 
 one can make similar considerations as in the previous section. Thus,
 \begin{equation}
  t^{(9)}_{6,22} \sim (-M_{22,9} ) \cdot e^{-{\rm i}\phi_1} \cdot M_{9,6} \cdot e^{{\rm i}\phi_2} \cdot (-M_{6,9} )  \cdot e^{-{\rm i}\phi_1} \cdot M_{9,22} . \label{eq:second_path}
 \end{equation}
 This amplitude describes a (a) propagation from the normal lead to the superconducting
 electrode $S1$ ($M_{9,22}$), (b)  Andreev reflection from electrode $S1$ ($e^{-{\rm i}\phi_1}$), (c) propagation of the hole-like state 
 from contact $S1$ to  $S2$ ($-M_{6,9}$), (d)  Andreev reflection of the hole-like particle on the contact $S2$ ($e^{{\rm i}\phi_2}$), (e) electron-like
 propagation between the superconducting electrodes $S1$ and $S2$ ($M_{9,6}$), 
 (f) a third Andreev reflection on the superconducting electrode $S1$ ($e^{-{\rm i}\phi_1}$), 
 (g) and a hole-like propagation from the contact $S1$ to the normal lead ($-M_{22,9}$). Finally, we also take into account 
 in our minimal model the amplitude $t_{9,22}$ describing the direct propagation between the normal lead
 and the lead $S1$ according to Eq.(\ref{eq:tkl}). The observed interference effect can be explained as the interplay between these two amplitudes:
 \begin{equation}
  I_N \sim |t_{9,22} + t^{(9)}_{6,22}|^2 = M_{9,22}^4 \cdot \left(1 + M_{9,6}^4 - 2 \cdot M_{9,6}^2 \cdot\cos (\phi_1-\phi_2 ) \right) \label{eq:unconv_current}
 \end{equation}
 The normal current $I_N$ in Eq(\ref{eq:unconv_current}) has  a maximum   at  phase difference $\phi_1-\phi_2=\pi$.   
The minus sign appearing  in front of the $\cos(\phi_1-\phi_2 )$ term in Eq(\ref{eq:unconv_current})
is due to   the peculiar properties of the Bogolioubov-de Gennes quasiparticles. 
 Namely, the amplitude $t^{(9)}_{6,22}$ contains one more hole-like propagation compared to the amplitude $t_{9,22}$, 
 which brings in an extra minus sign needed for the formation of the unconventional interference effect. 
 Note,  that this  argument  does not explain why the increase in $I_N$ appears only above a certain bias voltage.
 Moreover,  the transport process associated to the amplitude $t^{(9)}_{6,22}$ contains four more tunnelings between the 
 superconducting leads and the molecular core compared to the amplitude $t_{9,22}$.
 Thus, the amplitude $t_{9,22}$ might be expected to be much larger than the amplitude $t^{(9)}_{6,22}$ which would suppress the interference 
 effect between these two interfering paths.

The role of the ABSs  can be shown explicitly  by using  Green's function theory to calculate the differential 
conductance $\frac{{\rm d} I_N}{{\rm d}eV}$. 
{The details of the this calculation are  presented in the electronic supplementary information. For simplicity, let us assume that 
there is only one ABS in the system (the general case of more than one ABS is discussed in the electronic supplementary). 
Then the differential conductance can be approximated as\cite{Claughton_1995}
\begin{equation}
 \frac{{\rm d} I_N}{{\rm d}eV} \approx
 \frac{16e}{h}\frac{  \Gamma_{ABS,e}\Gamma_{ABS,h} }{(eV-E_{ABS})^2 + \Gamma_{ABS}^2}, 
 \label{eq:diff_cond_final_simplified_main_text}
\end{equation}
 where 
 $\Gamma_n=\Gamma_n^e+\Gamma_n^h$ is the level broadening of the ABS due to the presence of the normal lead, 
 $\Gamma_{ABS,e} = \langle ABS, e|W_N^{\dagger} \rm{Im}(g_N^e) W_N |ABS, e\rangle$ with $W_N$ being the coupling between 
 the normal lead and the central molecule, $g_N^e$ standing for the electron-like block of the surface
Green's function of the normal contact evaluated at energy $E_{ABS}$, and $|ABS, e\rangle$ represents the electron-like components of the wave function 
of the ABS. The definition of $\Gamma_{ABS,h}$ is analogous to  $\Gamma_{ABS,e}$ involving the hole-like
degrees of freedom instead of the electron-like components. 
According to Eq(\ref{eq:diff_cond_final_simplified_main_text}), the ABS leads to a resonant peak of Lorentzian lineshape 
in the differential conductance for $eV\approx E_{ABS}(\delta\phi)$.
The half-width of the resonance is  determined by the finite lifetime of the ABS which is due to the coupling to the normal lead 
given by $\Gamma_{ABS,e}$ and $\Gamma_{ABS,h}$. 
We note that a similar result can be obtained for a system hosting multiple ABSs. 
The total differential conductance in this case would be a sum of resonances centered on the energies of the individual ABSs'.
However, the ``cross-talk'' between the ABSs has an additional influence on the shape of the resonances, i.e.,  they start to deviate 
from the regular Lorentzian shape. (For details see the electronic supplementary material.)}

Looking back to Eq.~(\ref{eq:unconv_current}), one may now say that 
the interfering amplitude $t^{(9)}_{6,22}$ can be increased due to the Fabry-Perot-like resonant oscillations of the charged particles between the superconducting contacts.
These oscillations  lead to the formation of ABSs of finite lifetime, which, in turn, affect the current $I_N$ at  finite $eV$, as indicated by 
Eq.~(\ref{eq:diff_cond_final_simplified_main_text}).
\begin{figure}[h]
\begin{centering}
\includegraphics[scale=0.3]{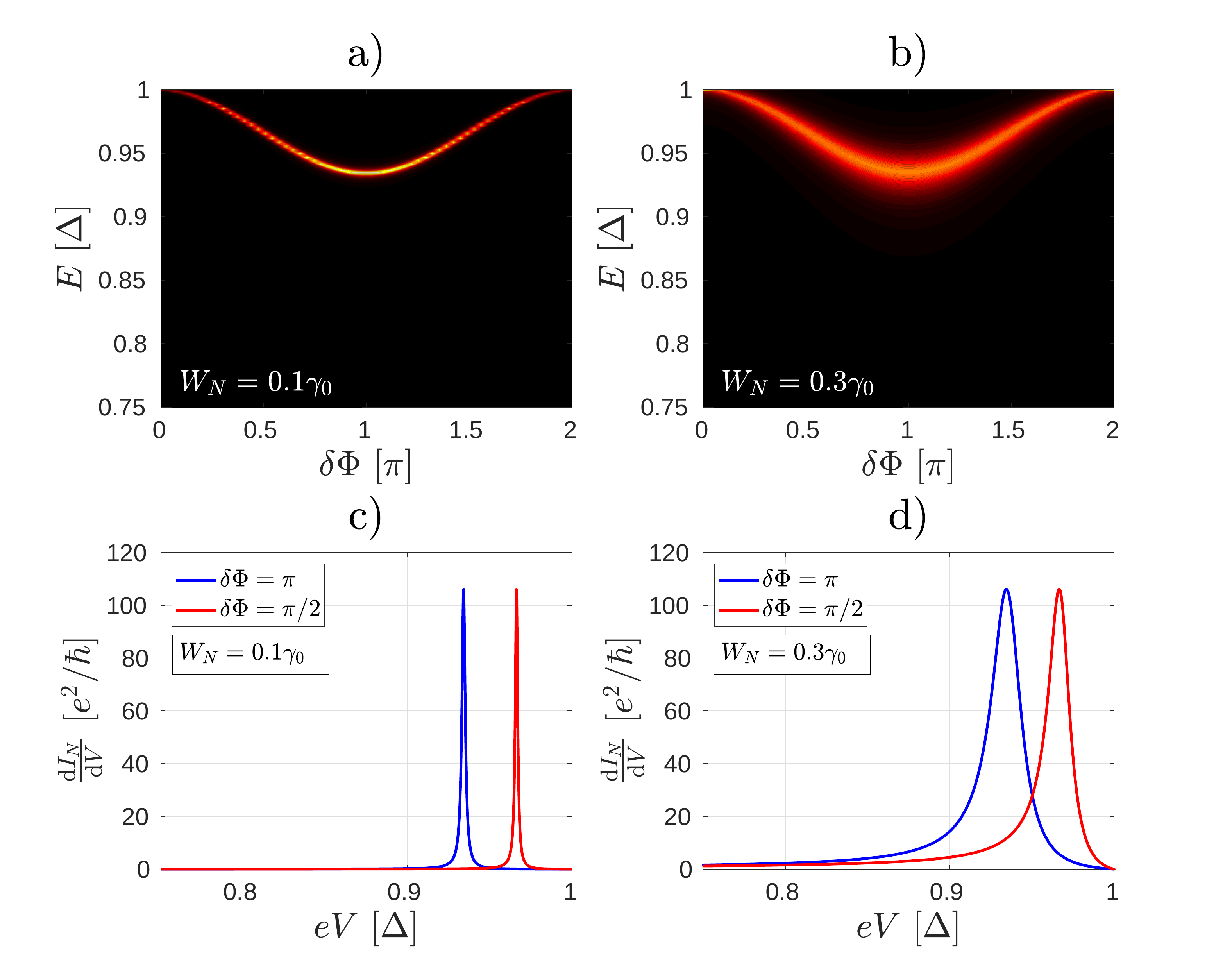}
\caption{(a) and (b): density of states of the molecular junction shown in Fig. \ref{fig:ABS-anthanthrene} for two different 
coupling $W_N$ of the normal lead to the molecule. The bright areas indicate  the dispersion of the ABS as a function of $\delta\Phi$.
The   ABS energy level  is broadened by increasing $W_N$. 
c) A resonance occurs in the differential conductance when the bias voltage $eV$ is close to the energy $E_n(\delta\phi)$ of the ABS in a).
d) As the ABS is broadened, the width of the resonance broadens as well.
}
\label{fig:diffcond_vs_gamma}
\end{centering}
\end{figure} 
The ABSs can be visualized by calculating the density of states of the junction (see the electronic
supplementary information for details). The results of such calculations for the system in Fig.~\ref{fig:ABS-anthanthrene} are shown 
in  Fig.~\ref{fig:diffcond_vs_gamma}. 
In  Figs.~\ref{fig:diffcond_vs_gamma}a) and b) we show the density of states for two different coupling $W_N$.  
The large values of the density of states (bright region) indicate the ABS. 
 In this particular case,  for $eV=0$ and zero temperature there is only one occupied ABS (at  energy $-E$, not shown) and one unoccupied ABS 
[at  energy $E$, Figs.\ref{fig:diffcond_vs_gamma}(a) and (b)].  
By applying a finite  $eV>0$ the occupation of these  ABSs can be changed, leading to the peculiar dependence of both $I_N$ and $I_s$ on $\delta\Phi$ 
in Fig.~\ref{fig:ABS-anthanthrene-currents} that we  noted  earlier. 
Because of the normal lead, the ABSs have a finite lifetime, which is determined  by the escape rate of the particles through the normal lead.
Therefore, the ABS lifetime (and consequently the width of the resonant peaks in the differential conductance) is expected to be sensitive 
to the coupling between the normal lead and the  central molecule. 
This can be clearly seen in Fig.~\ref{fig:diffcond_vs_gamma}c) and d),  where the peak of the differential conductance calculated 
for $W_N=0.1\gamma_0$ is considerably wider than the peak calculated for $W_N=0.3\gamma_0$.
{Notice, that for higher values of $W_N$ the resonant peak starts to deviate from the Lorentzian shape. 
This is because by increasing the coupling between the contacts and the central molecule one can no longer neglect the energy dependence of 
the Green's function of the normal contact in the calculation of  $\Gamma_{ABS,e}$ and $\Gamma_{ABS,h}$, see the electronic supplementary information.}
Since the ABS's energy $E_{ABS}$ depends on the phase difference $\delta\phi$, 
the peaks in $\frac{{\rm d} I_N}{{\rm d}eV}$ are also sensitive to the superconducting phase difference. 
This is also shown in Figs.~\ref{fig:diffcond_vs_gamma}(c) and (d). Therefore,  by measuring  
$\frac{{\rm d} I_N}{{\rm d}eV}$ as a function of  $\delta\phi$ 
one may obtain spectroscopic information about the ABSs\cite{vanWees-theory}. 

\begin{figure}[h]
\begin{centering}
\includegraphics[scale=0.3]{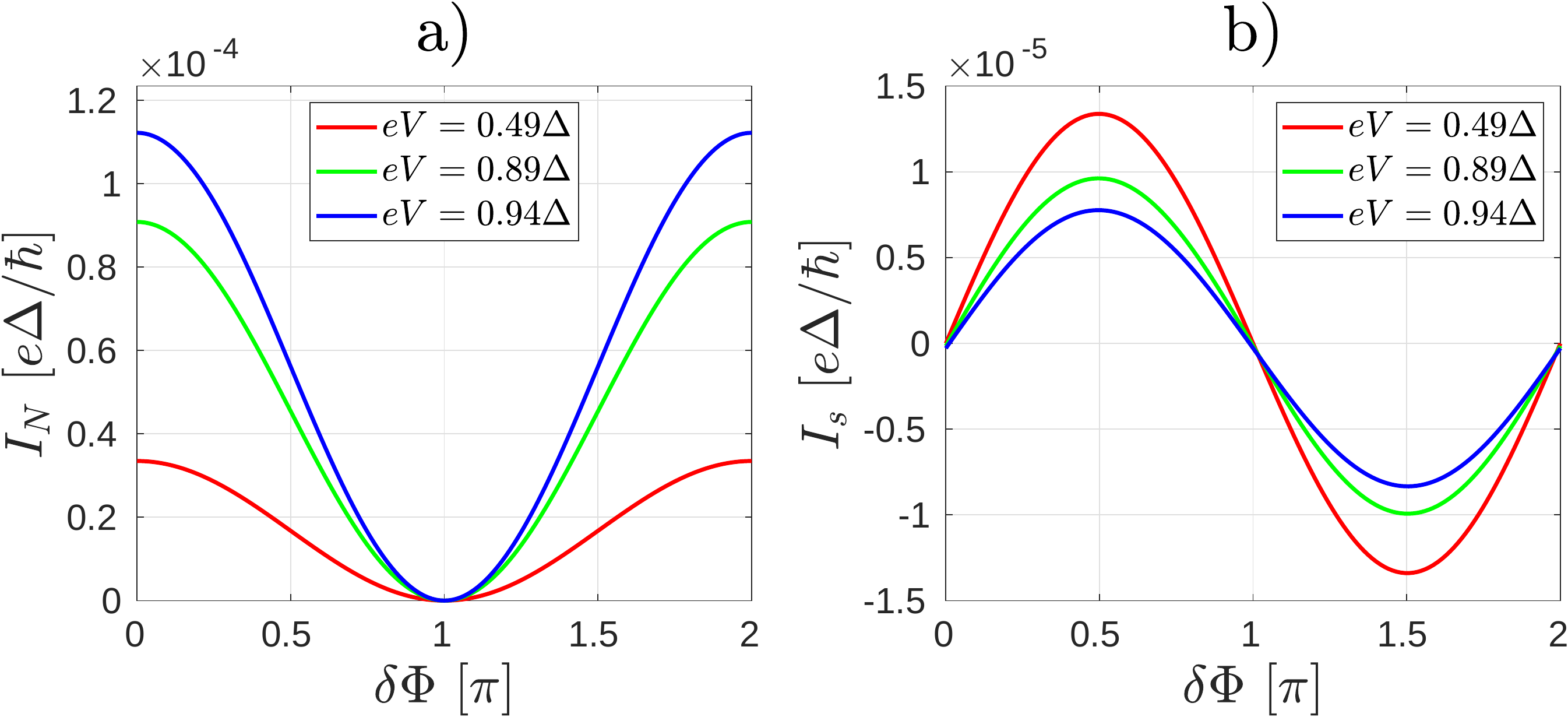}
\caption{The currents $I_N$ (a) and $I_s$ (b) as a function of the phase difference $\delta\phi$ between the superconducting leads 
for the system  depicted in Fig. \ref{fig:setup} for several bias voltage $eV$.  
a) the Andreev current $I_N$  shows a minimum at $\delta\phi=\pi$.  b) The current-phase dependence of supercurrent is $I_s \propto \sin\delta\phi$.
}
\label{fig:current-setup1}
\end{centering}
\end{figure}

The role of ABSs and  QI in the molecular core can be further illustrated by  studying the finite bias properties 
of the  system shown already in Fig.~\ref{fig:setup}, bottom panel. 
For this configuration of the leads the magic number vanishes between the two sites where 
the superconducting electrodes are attached. One may therefore expect that there is no ABS present in the system.  According to 
our calculations this is not exactly the case: one can find an ABS whose energy is very close to the value of the 
pair potential $\Delta$ in the leads, but it is nearly independent of $\delta\phi$ and therefore it can carry only a small supercurrent. 
This explains that for  a finite bias $eV$ the $\delta\phi$ 
dependence of $I_N$ remains qualitatively  the same  as in the  zero bias case discussed 
in Eqs(\ref{eq:tkl})-(\ref{eq:In-equilibrium}) and shows a minimum at 
$\delta\phi=\pi$ for all  bias voltages [Fig.~\ref{fig:current-setup1}(a)]. 
The supercurrent $I_s$ shows the conventional $\propto \sin\delta\phi$ dependence for $eV<\Delta$ [Fig.~\ref{fig:current-setup1}(b)].
By comparing Figs.\ref{fig:current-setup1}(a) and (b), one can see that although  a small $I_s$ can flow for 
finite $eV$,  the critical current $I_c$ is smaller than $I_N$. This is the opposite of what we found in the 
previous case [Fig.~\ref{fig:ABS-anthanthrene-currents}].
Overall, one may also notice that both $I_N$ and $I_c$ are much  smaller than previously, c.f. 
Figs.~\ref{fig:ABS-anthanthrene-currents} and \ref{fig:current-setup1}. 

These results  underpin the importance of ABSs in Andreev interferometers and are also a consequence of mid-gap transport. 
Namely,  the propagation amplitude described by the Green's function elements decay with the energy difference between the chemical potential 
and the energy of the eigenstates. Since the energy levels of the molecule are much further from the chemical potential than the ABS levels, 
their  contribution  to the Green's function elements will be also much smaller than the contribution of the ABSs. Thus, in the
mid-gap energy regime, the transport processes would indeed be dominated by the interference effects related to the ABSs.

\begin{figure}[h]
\begin{centering}
\includegraphics[trim={1.0cm 0.0cm 1.0cm 0.5cm},clip, width=1\linewidth]{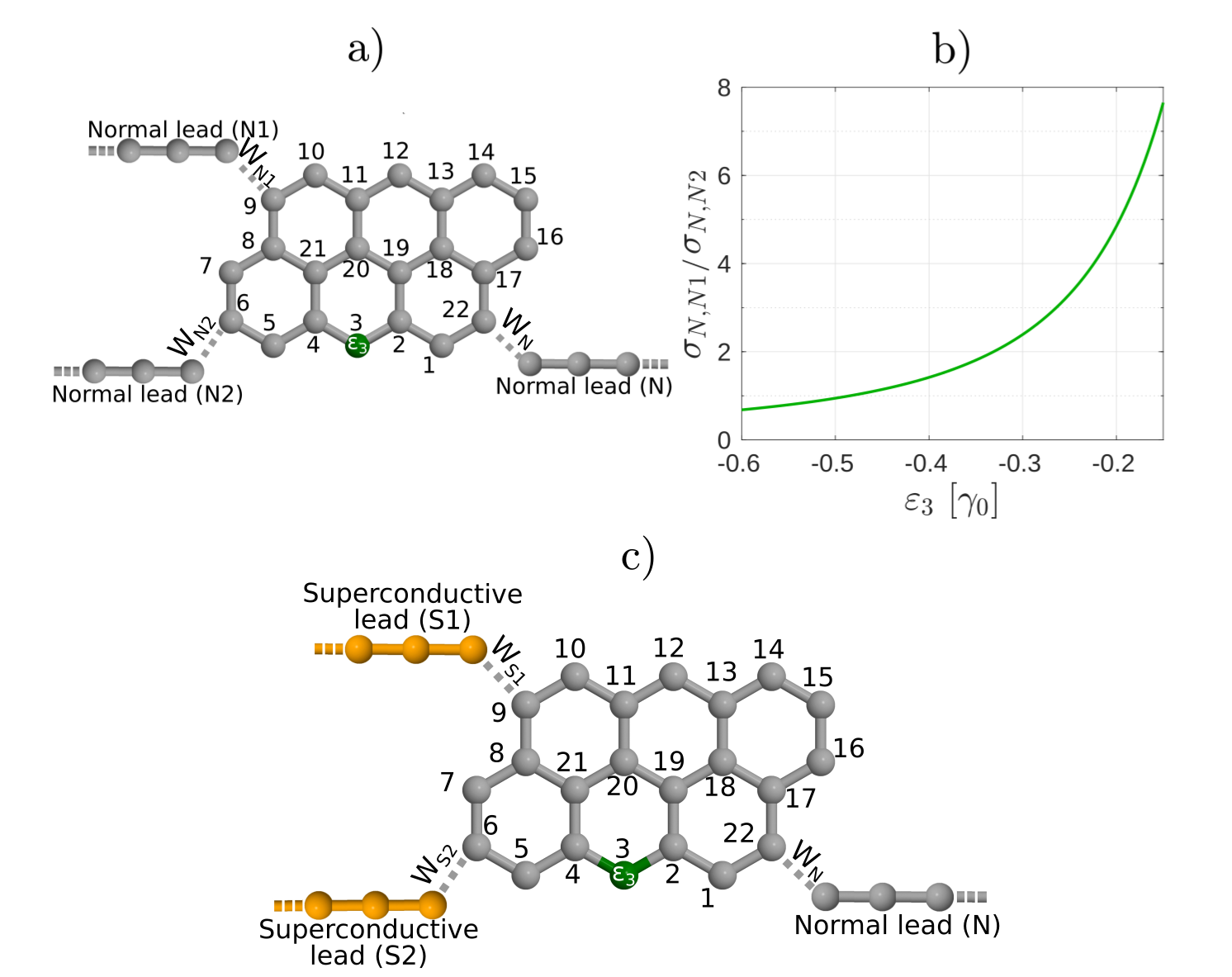}
\caption{a) Schematics of Anthanthrene molecule with a heteroatom denoted by green. 
b) Ratio of the normal conductance between contacts N-N2 and N-N1 as a function of the on-site energy  $\varepsilon_3$  
in Fig.~\ref{fig:heteroatom}.  
At $\varepsilon_3 = 0$ the conductance $\sigma_{N,N1}$ between contacts N and N1 is much larger than the conductance  $\sigma_{N,N2}$ 
between contacts N and N2,  in agreement with Refs.~\cite{Geng2015,Sangtarash2015}.  For finite $\varepsilon_3$ the conductance 
$\sigma_{N,N2}$  increases and can be comparable to $\sigma_{N,N1}$. c)  Andreev interferometer setup  obtained by replacing 
the normal leads $N1$ and $N2$ in a) by superconducting ones $S1$ and $S2$.}
\label{fig:heteroatom}
\end{centering}
\end{figure}

We now discuss the most general situation, where  ABSs can be found in the system and, in contrast with the case in 
Fig.~\ref{fig:setup}, the connectivity from the normal  lead to both superconducting terminal is finite.  
In what follows we shall  examine the unconventional interference effects as the connectivity in the interferometer is changed. 
We consider a setup similar to the one shown in Fig.~\ref{fig:setup} and  tune the asymmetry of the molecular interferometer  
by inserting  a substitutional heteroatom into the molecular core\cite{Sangtarash2016}, i.e. a carbon atom is replaced by a substituent heteroatom, as 
indicated schematically in  Fig.~\ref{fig:heteroatom}(a). 
Due to the presence of the heteroatom, new conductive channels open up in the molecular core that were originally closed 
via destructive QI effects. In our theoretical model we account for the presence of a substitutional heteroatom by a modified 
on-site energy on a  specific site in the molecule. By changing e.g., the on-site energy  $\varepsilon_3$ in the tight-binding Hamiltonian 
of the molecular core [see Fig.~\ref{fig:heteroatom}(a)], 
the normal conductance between sites labeled by even numbers also becomes finite\cite{Sangtarash2016}.
Assuming  that instead of S1 and S2 we have normal conducting leads N1 and N2 as in Fig.~\ref{fig:heteroatom}(a),  
the  evolution of the ratio of the zero-bias normal conductances $\sigma_{N,N1}$ and 
$\sigma_{N,N2}$ as a function of the on-site energy $\varepsilon_3$ is demonstrated in Fig.~\ref{fig:heteroatom}(b). 
As one can see, by varying  $\varepsilon_3$ one can gradually open a conductive channel between leads N and N2.

\begin{figure}[h]
\begin{centering}
\includegraphics[scale=0.3]{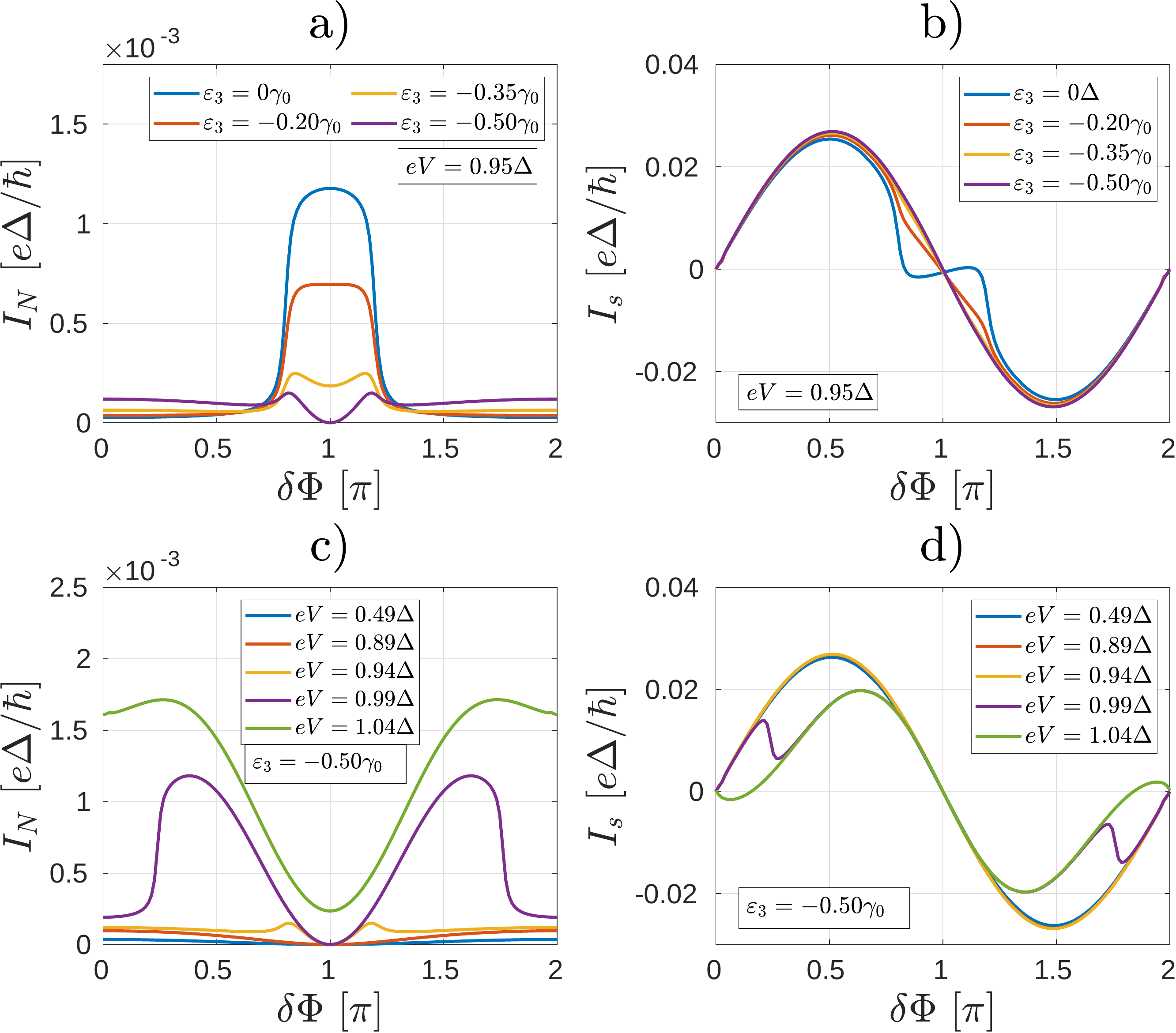}
\caption{The currents  (a)  $I_N$  and (b) $I_s$ (b) as a function of the phase difference $\delta\phi$ between the superconducting leads 
for the system  depicted in Fig. \ref{fig:heteroatom}(a) for several values of the   on-site energy $\varepsilon_3$ and 
fixed bias $eV=0.95\Delta$. $I_N$ starts to show a double peak structure as a function of   $\delta\phi$ for $\varepsilon_3 \gtrsim -0.20\gamma_0$. 
(c) $I_N$ and (d) $I_s$ as a function of the bias voltage $eV$ for $\varepsilon_3=-0.50\gamma_0$. 
}
\label{fig:heteroatom-currents}
\end{centering}
\end{figure}

We now consider  the finite bias properties of the Andreev interferometer shown in Fig.~\ref{fig:heteroatom}(c), which can be obtained be replacing 
the normal leads N1 and N2  by superconducting ones S1 and S2 in Fig.~\ref{fig:heteroatom}(a). 
First, we calculate $I_N$ for several values of $\varepsilon_3$ and fixed $eV=0.95\Delta$. 
Remarkably, as shown in Fig.~\ref{fig:heteroatom-currents}(a),  $I_N$ takes on a hat-like shape with two maxima around 
phase the difference $\pi$  for such values of $\varepsilon_3$, where  $\sigma_{N,N1}$  and $\sigma_{N,N2}$ are comparable. 
This is clearly different from the results in Fig.~\ref{fig:ABS-anthanthrene-currents}(a) and we are not aware of  similar 
results in mesoscopic NS systems. 
Regarding  $I_s$ [Fig.~\ref{fig:heteroatom-currents}(b)], for smaller values of $\varepsilon_3$ where $\sigma_{N,N1}\gg\sigma_{N,N2}$, 
it is  qualitatively  similar to the results shown in Fig.~\ref{fig:ABS-anthanthrene-currents}(b).
On the other hand,  the current-phase relation of $I_s$  becomes similar to the conventional $I_s=I_c\sin\delta\phi$ as the conductive 
channel gradually opens between N and S2 and consequently $\sigma_{N,N1} \approx \sigma_{N,N2}$
[see e.g., the case $\varepsilon_3=-0.50 \gamma_0$ in Fig.~\ref{fig:heteroatom-currents}(b)]. Note that in this case the $I_s(\delta\phi)$ dependence for  
$\delta\phi\approx \pi$ is  different from the corresponding $eV=0.94\Delta$ result shown in Fig.~\ref{fig:ABS-anthanthrene-currents}(b).
In Fig.~\ref{fig:heteroatom-currents}(c) and (d) we show $I_N$ and $I_s$, respectively, as a function of $\delta\phi$ 
for several biases $eV$.  Here we fixed $\varepsilon_3=-0.50\gamma_0$, i.e., $\sigma_{N,N1}\approx\sigma_{N,N2}$. 
As one can see, for small $eV$, when the occupation of the ABS is not yet modified, $I_N$ shows qualitatively the same behavior 
as in Fig.\ref{fig:current-setup1}(a), i.e., when there was no current-carrying ABS in the system. For larger $eV$, however, 
there is a clear difference with respect to both Fig.~\ref{fig:ABS-anthanthrene-currents}(a)  and Fig.~\ref{fig:current-setup1}(a), 
since $I_N$ adopts a hat-like dependence on $\delta\phi$. The non-equilibrium population of the ABSs  
also affects $I_s$ [see Fig.~\ref{fig:heteroatom-currents}(d)] which starts to deviate from the $\propto\sin\delta\phi$ dependence 
for $eV>0.95\Delta$.

\begin{figure}[h]
\begin{centering}
\includegraphics[scale=0.3]{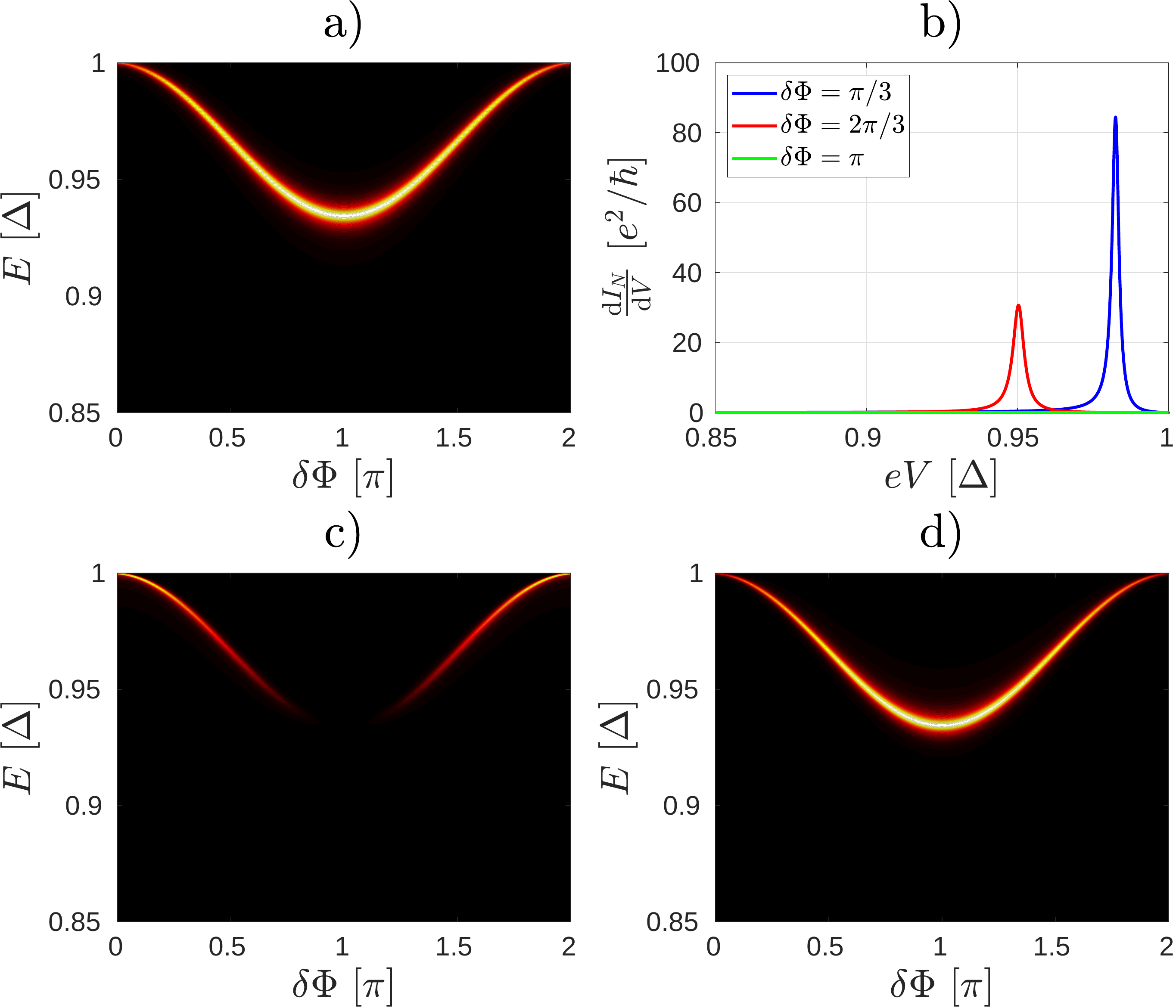}
\caption{a) The density of states of the ABS for the case shown in Fig.~\ref{fig:heteroatom}(c). b) The differential conductance 
corresponding to a).  The local density of states for  electron (c)  and  hole (d) quasiparticles as a function of $\delta\Phi$. 
In these calculations we used $\epsilon_3=-0.50\gamma_0$.
}
\label{fig:heteroatom-DOS}
\end{centering}
\end{figure}

According to our calculations the presence of a heteroatom does not modify the ABS spectrum significantly [Fig.\ref{fig:heteroatom-DOS}(a)]. 
As shown in Fig.~\ref{fig:heteroatom-DOS}(b), 
when $I_N$ nearly vanishes for $\varepsilon_3=-0.50\gamma_0$, $\delta\phi=\pi$  
[Fig.~\ref{fig:heteroatom-currents}(a)], the differential  conductance  $\frac{{\rm d} I_N}{{\rm d}eV}$ vanishes, too.  
According to Eq.~(\ref{eq:diff_cond_final_simplified_main_text}), the vanishing of $\frac{{\rm d} I_N}{{\rm d}eV}$ can be explained 
only if the coupling between the normal lead and the ABS vanishes. 
Therefore we turn our attention to the electron- and hole-like broadening terms $\Gamma^e_n$ and $\Gamma^h_n$ in the numerator of Eq.~(\ref{eq:diff_cond_final_simplified_main_text}). In Figs\ref{fig:heteroatom-DOS}(c) and (d) we show   the local density 
of states (LDOS) on the molecular site connected to the normal contact separately for the electron- and hole-like degrees of freedom. 
Note that   $\Gamma^e_n$ and $\Gamma^h_n$ are proportional to the corresponding LDOS. 
As one can see,  the electron-like component of the LDOS becomes highly suppressed at phase difference $\delta\phi= \pi$, 
while the hole-like components has a maximum there. 
{In turn, we found that on other sites of the molecule the  hole-like component of the LDOS can be suppressed and the electron-like LDOS enhanced 
 (an example is shown in the Supplementary information). }

\begin{figure}[h]
\begin{centering}
\includegraphics[trim={0cm 0cm 0cm 0cm},clip, width=0.9\linewidth]{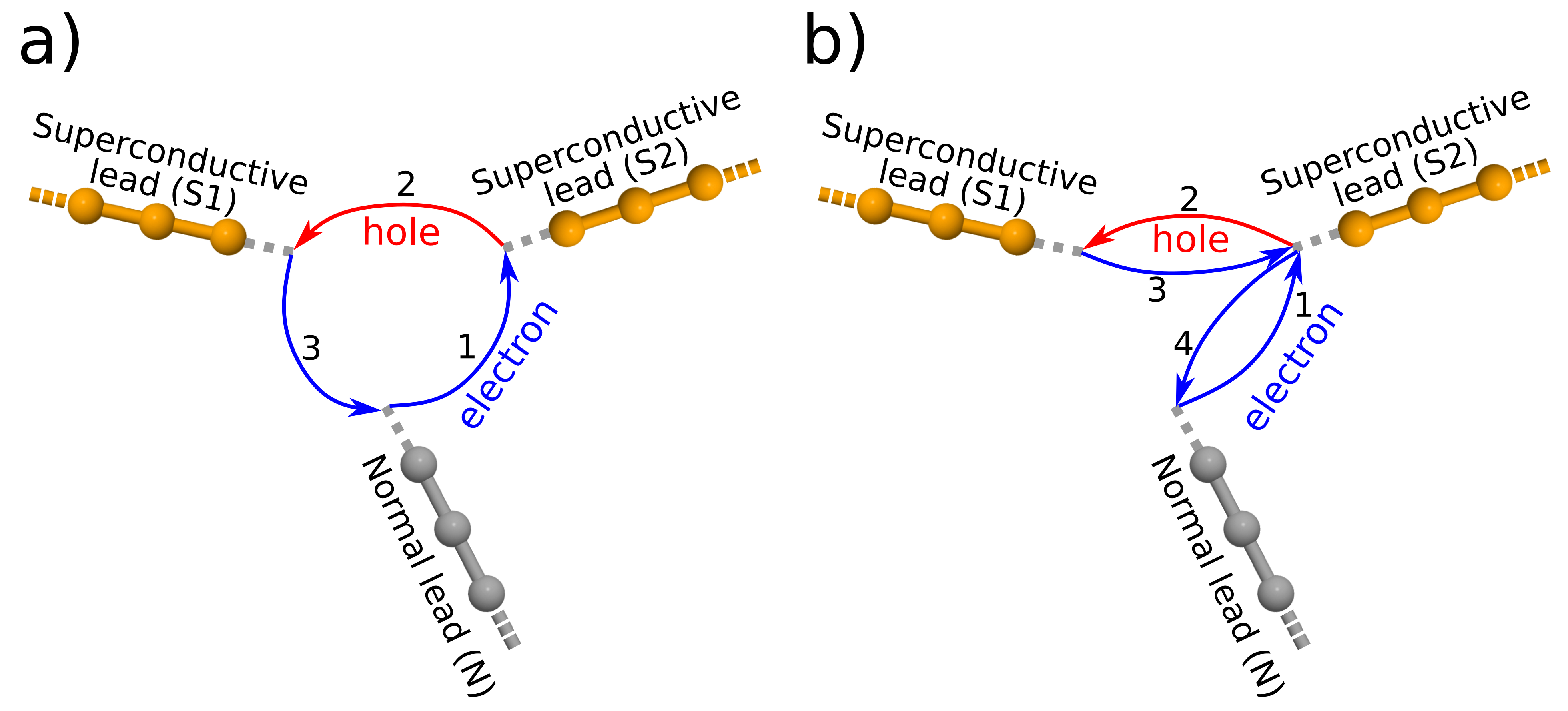}
\caption{An example for  interfering paths having an amplitude of opposite sign for the electron- and hole-like particles. 
These kinds of paths have an odd number of propagation through the molecular core. b) A representative of trajectories consisting 
of even number of propagations through the molecular core. The amplitude of these kinds of trajectories
have the same sign for the electron- and hole-like quasiparticles.}
\label{fig:heteroatom-trajectories}
\end{centering}
\end{figure}

The surprising result in Figs.\ref{fig:heteroatom-DOS}(c) and (d) can be understood as a peculiar interference effect that acts in a different way 
on  the electron- and the hole-like particles. (Note, that due to  electron-hole symmetry, the same feature can be observed 
for negative energies with a constructive interference 
in the electron-like part of the LDOS and with a destructive interference in the hole-like part of the LDOS.)  
One can give the following simple argument  in terms of new quasiparticle paths. 
In  Fig.~\ref{fig:heteroatom-trajectories} we show two quasiparticle trajectories.  The first describes the process  
N $\rightarrow$ S2 $\rightarrow$ S1 $\rightarrow$ N  and the last segment S1 $\rightarrow$ N is made possible by the fact that the 
substitutional heteroatom opened a new conductive channel.
Using a similar argument as in the case of Eq.~(\ref{eq:second_path}), one can argue that  the amplitude 
of the path contributing to the electron-like part of the wave function can be expressed as
\begin{eqnarray}
t_{\rm odd}^e  &\sim& M_{N,S1}\cdot e^{i\phi_1}\cdot (-M_{S1,S2}) \cdot e^{-i\phi_2} \cdot M_{S2,N}\nonumber\\
&\sim& -M_{N,S1}\cdot M_{S1,S2}\cdot M_{S2,N} e^{i(\phi_1-\phi_2 )}.
\label{eq:t_e_odd}
\end{eqnarray}
Since $t_{\rm odd}^e$ contains an odd number of propagations through the molecule, and the sign of the propagation depends on 
whether one considers electron- or hole-like particles, the amplitude $t_{\rm odd}^h$ contributing to the hole-like component of the wave function
would differ by a minus sign compared to $t_{\rm odd}^e$. 
Now consider the process N $\rightarrow$ S2 $\rightarrow$ S1 $\rightarrow$ S2 $\Rightarrow$ N depicted 
in Fig.~\ref{fig:heteroatom-trajectories}(b). The last propagation indicated by S2 $\Rightarrow$ N describes 
a normal reflection (without electron-hole conversion) at the site connected to S2  and a forthcoming propagation
to the normal contact. (Since we are working in the weak coupling limit, the normal reflection at sites connected to the contacts
has a finite probability.) 
The amplitude corresponding to this path can be expressed as follows:
\begin{eqnarray}
t_{\rm even}^e  &\sim & M_{N,S2}\cdot M_{S2,S1} \cdot e^{i\phi_1}\cdot (-M_{S1,S2}) \cdot e^{-i\phi_2} \cdot M_{S2,N}\nonumber\\
& \sim & -M_{N,S2}^2 \cdot M_{S1,S2}^2 \cdot e^{i(\phi_1-\phi_2 )}
\label{eq:t_e_even}
\end{eqnarray}
Since $t_{\rm even}^e $ depends on the square of the connectivity matrix elements, the corresponding hole-like amplitude 
$t_{\rm even}^h $ would have the same sign as $t_{\rm even}^e$. 
One can see that because of the sign difference, 
there is a destructive interference in total amplitude  $t^e=t_{\rm even}^e+t_{\rm odd}^e$ and a constructive one in 
 $t^h=t_{\rm even}^h+t_{\rm odd}^h$. This example shows how differences can appear in the processes that determine the electron-like and 
 the hole-like LDOS. Note  that, strictly speaking,  in the calculation of  
 $t_{}^e$ and $t^h$  one would need to take into account all possible scattering  paths and not only those 
discussed above.  We expect, however, that the described interference effect would not be affected significantly. 

We also mention that according to our numerical results the interference effect can be swapped between the electron- 
and hole-like components by changing the sign of the on-site energy
$\varepsilon_3$ of the heteroatom. According Eq. (8) of Ref.\cite{Sangtarash2016} the connectivity
matrix element $M_{N,S1}$ can change a sign for sufficiently large heteroatom on-site energy. Consequently, $t_{odd}^e$ would 
also change  sign resulting in a constructive interference for the electron-like and destructive interference for the hole-like 
components in the  LDOS.

\begin{figure}[h]
\begin{centering}
\includegraphics[trim={0cm 0cm 0cm 0cm},clip, width=1\linewidth]{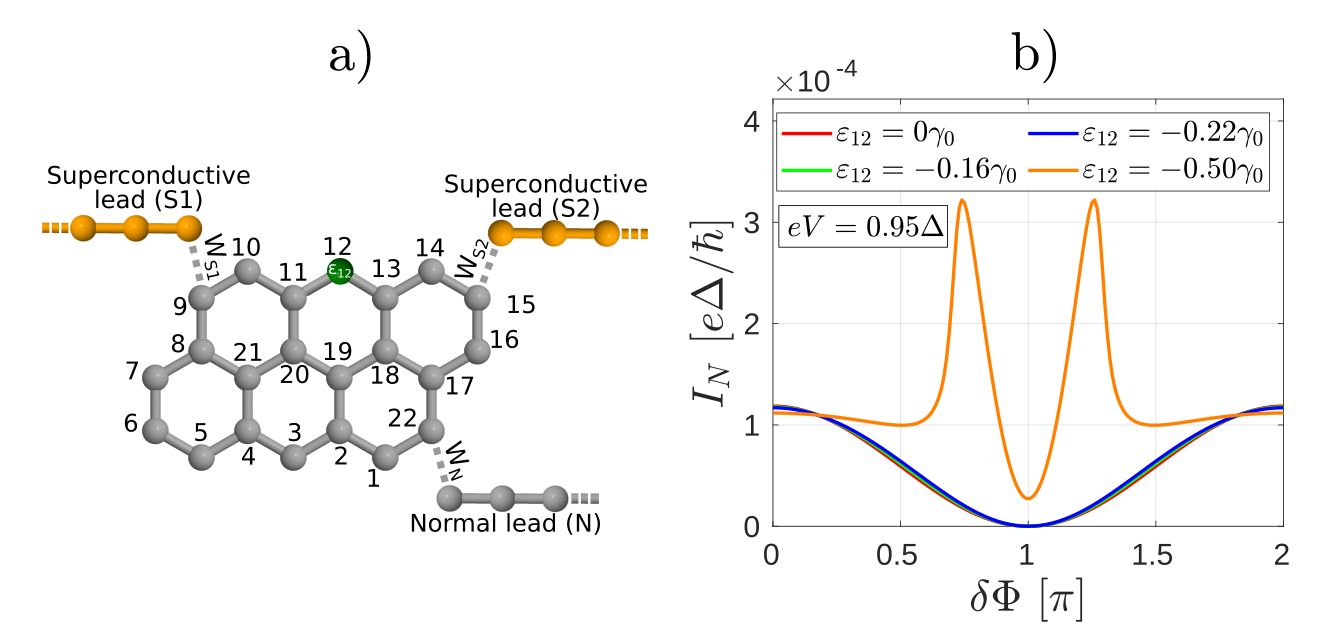}
\caption{a) Anthanthrene molecule with a heteroatom (denoted by green) and the same configuration of leads as in Fig.~\ref{fig:setup}. 
b) The normal current $I_N$  as a function of the phase  difference $\delta\Phi$  between the superconducting leads for the setup in (a). }
\label{fig:heteroatom-between-supraleads}
\end{centering}
\end{figure}

Opening of new conductance channels can affect the properties of the molecular Andreev interferometer not only in the case 
discussed in Figs.~\ref{fig:heteroatom} and \ref{fig:heteroatom-currents}, where the conductance between the leads $N$ and $S1$ was 
tuned. As noted earlier for the system in Fig.~\ref{fig:setup},  for this configuration of the leads the
connectivity matrix element is zero between the two sites where the superconducting electrodes are attached. 
However, this  connectivity matrix element can also be made  finite by adding a heteroatom  as indicated in  Fig.~\ref{fig:heteroatom-between-supraleads}(a). 
This means  changing the onsite energy  $\epsilon_{12}$ in the tight-binding Hamiltonian of the molecular core.
We found that the dependence of the supercurrent on  $\epsilon_{12}$ and 
$eV$ is qualitatively similar to the behavior in Fig.~\ref{fig:heteroatom-currents}(b) and (c). Therefore we only show the calculations for $I_N$
in Fig.~\ref{fig:heteroatom-between-supraleads}(b). As the connectivity grows for larger values of  $\epsilon_{12}$, the $\delta\Phi$ dependence 
of $I_N$ also undergoes a drastic change and, interestingly, adopts a qualitatively similar behavior to the one 
shown in Fig.~\ref{fig:heteroatom-currents}(a), i.e., there are two maxima in the current around $\delta\Phi=\pi$. 


\section*{Conclusions and Outlook}

In this article, we have investigated the interplay between two quantum interference phenomena that take place on hugely different energy scales;
QI within molecules, which takes place on the scale of electron volts and QI associated with superconductivity, which takes place on 
the scale of milli-electron volts. 
We studied the interplay between connectivity-driven QI in molecular cores and non-equilibrium charge distribution in 
three-terminal Andreev interferometers based on PAH molecules. 
We showed that  QI determines certain fundamental properties of the ABS in the system, while their energies can be tuned by the phase 
difference between the superconducting probes. 
Consequently, QI and the non-equilibrium ABS occupation in the molecular core, which can be modified by a bias voltage applied to
the normal lead, affects both the normal current  and the supercurrent in the system. 
We gave a simplified explanation  of some of the complicated interference effects in terms of electron and hole trajectories  
and point out when such explanation breaks  down under non-equilibrium conditions. 
We found  that the dependence of the normal current on the  superconducting phase difference can exhibit a double-peak  structure,
while the supercurrent can show a $\pi$ transition when the bias  $eV$ on the normal lead is larger than the superconducting gap.  
We  also showed that adding a heteroatom to the PAH core can significantly change the QI and can induce an asymmetry in the spatial distribution 
of the electron- and hole-like particles, which has a direct impact on the phase dependence of the normal current. This indicates that the 
properties of molecular Andreev interferometers can be tuned by engineering QI in the molecular core. 

{For the future one may  envisage  a system similar to the one shown in Fig.~\ref{fig:heteroatom} but with two normal leads (N1 and N2) attached to different sites of the 
molecular core. Assume now that lead N1 would be coupled  to a site where, e.g.,  the electron LDOS is enhanced  and the hole suppressed, whereas lead N2 to a site where 
the opposite is true,  i.e., the electron LDOS is suppressed and the hole LDOS is enhanced.  Then the so-called non-local Andreev reflection (N1 $\rightarrow$ N2), 
where an incoming electron from lead N1 is Andreev reflected into lead N2,  would be enhanced with respect to  local Andreev reflection (N1 $\rightarrow$ N1) 
and normal electron transmission (N1 $\rightarrow$ N2). Therefore,  in such  four-terminal device 
the  asymmetry between the electron- and hole-like degrees of freedom on certain sites of the molecular core could be translated 
into a spatial separation of electron pairs originating from  the superconducting condensate. This process is  called Cooper pair splitting and it provides 
entangled electron pairs that may play an important role in quantum information processing. Most of  the proposed Cooper pair splitters  to-date relied on
Coulomb blockade transport through quantum dots\cite{firstCPS,Hofstetter2009,PhysRevLett.104.026801, PhysRevLett.107.136801}, 
or on peculiar properties of novel low-dimensional materials\cite{PhysRevLett.100.147001,PhysRevB.84.115420,PhysRevB.92.241404}.  
Our results indicate that Cooper pair splitting  may also be achieved in multi-terminal molecular systems where the spatial separation of the
Cooper pairs would rely  on the inner QI effects of the  molecule. The detailed study of such four-terminal molecular Cooper pair splitters is an 
interesting problem which we   leave  to a future work. }

\section*{Acknowledgements}
N.L.P, P.R.,  A.K. and J. Cs. were supported by NKFIH within the Quantum Technology National Excellence Program 
(Project No. 2017-1.2.1-NKP-2017-00001) and by Hungarian Scientific Research Fund, OTKA Grant No. NN 127903 (Topograph FlagERA
project).  P. R. also acknowledges the funding from OTKA PD123927 and K123894 and N.L.P was also supported by the ÚNKP-19-1 New National Excellence Program of the Ministry for Innovation and Technology. C.J.L. acknowledges financial support from the UK EPSRC, through grant nos. EP/M014452/1, EP/P027156/1 and EP/N03337X/1. This work was additionally supported by the European Commission is provided by the FET Open project 767187 – QuIET.

\appendix

\section{Theoretical background to calculate the differential conductance}
\label{sec:perturb_expansion}

In this section we give the technical details to calculate and analyze the differential conductance on the normal lead connected to an Andreev interferometer.
The aim of this section is twofold.
Firstly we obtain a closed formula which can be evaluated numerically.
Secondly, we answer the questions raised in the discussion of the results  in Fig.3 of the main text.
Namely, the reported unconventional interference effect is manifested only above a certain bias voltage applied on the normal lead.
Secondly, the amplitude of the interfering path $N\rightarrow \textrm{mol}\rightarrow S2\rightarrow \textrm{mol} \rightarrow S1\rightarrow \textrm{mol} \rightarrow S2 \rightarrow \textrm{mol}\rightarrow N$ depicted in Fig.~3. of the main text is expected to be much smaller than the amplitude of the interfering path $N\rightarrow \textrm{mol}\rightarrow S2\rightarrow \textrm{mol} \rightarrow N$, yet the resulting interference pattern in the Andreev current seem to be quite robust (see Fig.~4 in the main text).
(The interfering path $N\rightarrow \textrm{mol}\rightarrow S2\rightarrow \textrm{mol} \rightarrow S1\rightarrow \textrm{mol} \rightarrow S2 \rightarrow \textrm{mol}\rightarrow N$ depicted in Fig.~3. of the main text involves four extra tunnelings between the leads and the central molecule compared to the interfering path $N\rightarrow \textrm{mol}\rightarrow S2\rightarrow \textrm{mol} \rightarrow N$.)

The Andreev current can be evaluated using Eq.~(4) of the main text.
In this equation the lesser Green's function $G^<$ can be calculated within the Keldysh non-equilibrium framework using the Keldysh equation\cite{PhysRevB.71.024517,PhysRevB.70.104511,PhysRevB.68.075306,NEGFtheory,Pala.2007}:
\begin{equation}
 G^{<} = G^R\Sigma^<G^{A}, \label{eq:Keldysh_kinetic}
\end{equation}
where $G^R(E)$ [$G^A(E)$] is the retarded [advanced] Green's function and $\Sigma^<(E) = \Sigma^<_{S1}(E) + \Sigma^<_{S2}(E) + \Sigma^<_{N}(E,V)$ contains the lesser self energies of the leads.

The differential conductance can be derived from Eq.~(4) of the main text utilizing the relation given by Eq.~(\ref{eq:Keldysh_kinetic}):
\begin{widetext}
\begin{equation}
 \frac{{\rm d} I_N}{{\rm d}eV} = -\frac{2e}{h}{\rm Re}\left\{\frac{{\rm d}}{{\rm d}eV}\int dE \;{\rm Tr} \left[ \tau_3W_N G^R\bigg(\Sigma^<_{S1} + \Sigma^<_{S2} + \Sigma^<_{N}(eV)\bigg)G^{A}\right ]\right\}
\end{equation}
\end{widetext}
This expression can be further simplified by applying the derivation with respect to the bias voltage $V$ on the integrand. 
Notice that only the self energy of the normal lead depends on $eV$. Hence
\begin{equation}
 \frac{{\rm d} I_N}{{\rm d}eV} = -\frac{2e}{h}{\rm Re}\left\{\int dE \;{\rm Tr} \left[ \tau_3W_N G^R\frac{{\rm d}}{{\rm d}eV}\Sigma^<_{N}(E,eV)G^{A}\right ]\right\} \label{eq:diff_integral}
\end{equation}
Furthermore, the lesser self energy $\Sigma^<_{N}(E,eV)$ depends on the bias voltage via the thermal occupation number. 
In the electron-hole space the lesser self energy can be given as\cite{PhysRevB.68.075306}
\begin{widetext}
\begin{eqnarray}
 \Sigma^<_{N} &=& 
 \begin{pmatrix}
  f_e (\Sigma^R_{N,e} - \Sigma^A_{N,e}) & 0 \\ 0 & -f_h(\Sigma^R_{N,h} - \Sigma^A_{N,h})
 \end{pmatrix} \\ &=&
 \begin{pmatrix}
  f_e \bigg(\left(g^{A}_{N,e}\right)^{-1} - \left(g^R_{N,e}\right)^{-1}\bigg) & 0 \\ 0 & f_h\bigg(\left(g^{A}_{N,h}\right)^{-1} - \left(g^R_{N,h}\right)^{-1}\bigg)
 \end{pmatrix},  \label{eq:self_energy}
\end{eqnarray}
\end{widetext}
where $f_e=f(E-eV)$ [$f_h=f(E+eV)$] is the thermal occupation number for the electrons [holes] given by the Fermi-distribution 
function and $\Sigma^R_{N,e}$ [$\Sigma^A_{N,e}$] and $\Sigma^R_{N,h}$ [$\Sigma^A_{N,h}$] are the retarded [advanced] self energies of the 
electron-like and hole-like particles in the normal lead, uncoupled from the rest of the system.
Similarly, $g^R_{N,e/h}$ and $g^A_{N,e/h}$ stand for the retarded and advanced Green's functions of the electron/hole-like particles in the normal lead.
To calculate the retarded and advanced self energies and Green's functions we followed the numerical procedure described in Ref.~\cite{PhysRevB.78.035407}.
Also, we assume the uncoupled leads to be in thermal equilibrium.

For simplicity we will consider the zero temperature limit in our calculations. 
Consequently, the derivative of the Fermi distribution function is the Dirac delta function and the integral in 
Eq.~(\ref{eq:diff_integral}) simplifies to
\begin{widetext}
\begin{equation} 
\begin{split}
  \frac{{\rm d} I_N}{{\rm d}V} = -\frac{2e}{h}{\rm Re}\left\{{\rm Tr} \left[ \tau_3W_N G^R(eV)\begin{pmatrix}
  \left(g^{A}_{N,e}(eV)\right)^{-1} - \left(g^R_{N,e}(eV)\right)^{-1} & 0 \\ 0 & 0
 \end{pmatrix}G^A(eV)\right ] \right\}   \\ 
 + \frac{2e}{h}{\rm Re}\left\{{\rm Tr} \left[ \tau_3W_N G^R(-eV)\begin{pmatrix}
  0 & 0 \\ 0 & \left(g^{A}_{N,h}(-eV)\right)^{-1} - \left(g^R_{N,h}(-eV)\right)^{-1}
 \end{pmatrix}G^A(-eV)\right ]\right\} \label{eq:diff_cond}
\end{split}.
\end{equation}
\end{widetext}
As we can see from Eq.~(\ref{eq:diff_cond}), the key element to calculate the differential conductance is the retarded and 
advanced Green's functions $G^R$ and $G^A$. 
Eq.~(\ref{eq:diff_cond}) then can be directly used to calculate numerically the differential conductance in the studied three-terminal junctions.

To get further insight into the physics of the transport process we follow the logic of Ref.~\cite{Claughton_1995} to evaluate these Green's functions 
in terms of the Dyson's equation.
Let us denote the retarded Green's function of the unified system of the two superconducting contacts and the central molecular core by $g^R_{\textrm{mol}}$.
Then the retarded Green's function of the whole Andreev interferometer can be evaluated in terms of the Dyson's equation:
\begin{equation}
\label{eq:Dyson_eq}
 G^R =  \left(\begin{array}{cc} \left(g^R_{\textrm{mol}}\right)^{-1} & -W_N^{\dagger} \\ -W_N & \left(g^R_{N}\right)^{-1} \end{array}\right)^{-1},
\end{equation}
where 
\begin{equation}
 g^R_N =  \left(\begin{array}{cc} g^R_{N,e} & 0\\ 0 & g^R_{N,h} \end{array}\right)
\end{equation}
is the Green's function of the normal lead containing both the electron and hole-like components.
Equation (\ref{eq:Dyson_eq}) yields for the individual components of the Green's function:
\begin{widetext}
\begin{equation}
 G^R =  \left(\begin{array}{cc} G^R_{\textrm{mol,mol}} & G^R_{\textrm{mol,N}} \\ G^R_{\textrm{N,mol}} & G^R_{\textrm{N,N}} \end{array}\right) = \left(\begin{array}{cc} g^R_{\textrm{mol}}\left(1-W_N^{\dagger} g^R_N W_N g^R_{\textrm{mol}}\right)^{-1} & g^R_{\textrm{mol}}\left(1-W_N^{\dagger} g^R_N W_N g^R_{\textrm{mol}}\right)^{-1}W_N^{\dagger} g^R_N \\ g^R_N \left(1-W_N g^R_{\textrm{mol}} W_N^{\dagger} g^R_N \right)^{-1}W_N g^R_{\textrm{mol}} & g^R_N\left(1-W_N g^R_{\textrm{mol}} W_N^{\dagger} g^R_N \right)^{-1} \end{array}\right). \label{eq:evaluated_Dyson}
\end{equation}
\end{widetext}
Considering the rules of the matrix multiplication, and that the only non-zero elements of the lesser self energy of Eq.~(\ref{eq:self_energy}) are the block diagonal parts related to the leads, in order to evaluate the differential conductance (\ref{eq:diff_cond}) it is enough to consider the $G^R_{\textrm{mol},N}$ block of the retarded Green's function and the $G^A_{N,N}$ part of the advanced Green's function.
According to the structure of Eq.~(\ref{eq:evaluated_Dyson}) one finds:
\begin{widetext}
\begin{equation}
\begin{split}
 G^R_{N,N} &= g^R_N \sum_{n=0}^{\infty} \left(W_N g^R_{\textrm{mol}} W_N^{\dagger} g^R_N \right)^n = g^R_N + g^R_N W_N g^R_{\textrm{mol}}\sum_{n=0}^{\infty} \left(W_N g^R_{\textrm{mol}} W_N^{\dagger} g^R_N \right)^nW_N^{\dagger} g^R_N  \\ 
 &= g^R_N + g^R_N W_N g^R_{\textrm{mol}} \left(1-W_N^{\dagger} g^R_N W_N g^R_{\textrm{mol}} \right)^{-1} W_N^{\dagger} g^R_N = g^R_N +g^R_N W_N G^R_{\textrm{mol,mol}} W_N^{\dagger} g^R_N,
\end{split} \label{eq:G_N_N}
\end{equation}
\end{widetext}
and
\begin{equation}
 G^R_{\textrm{mol},N} = G^R_{\textrm{mol,mol}} W_N^{\dagger} g^R_N\;. \label{eq:G_mol_N}
\end{equation}
We now return to the evaluation of the differential conductance given by Eq.~(\ref{eq:diff_cond}).
For simplicity we continue our calculations focusing on the first (electron-like) part of Eq. (\ref{eq:diff_cond}).
(Due to the electron-hole symmetry of the Bogoliubov-de Gennes equations, the hole-like part would give the same result.)
Inserting Eqs.~(\ref{eq:G_N_N}) and (\ref{eq:G_mol_N}) into Eq.~(\ref{eq:diff_cond}) yields:
\begin{widetext}
\begin{equation} 
\begin{split}
  \frac{{\rm d} I^e_N}{{\rm d}V} &= -\frac{2e}{h}{\rm Re}\left\{{\rm Tr} \left[ \tau_3W_N G^R_{\textrm{mol},N}\begin{pmatrix}
  \left(g^{A}_{N,e}\right)^{-1} - \left(g^R_{N,e}\right)^{-1} & 0 \\ 0 & 0
 \end{pmatrix}G^A_{N,N}\right ] \right\}  \\
  &= -\frac{4e}{h}{\rm Im}\left\{{\rm Tr} \left[ \tau_3W_N G^R_{\textrm{mol,mol}} W_N^{\dagger} \begin{pmatrix}
  {\rm Im}\left( g^{R}_{N,e}\right) & 0 \\ 0 & 0
 \end{pmatrix}\left(1 + W_N G^A_{\textrm{mol,mol}} W_N^{\dagger} g^A_N\right)\right ] \right\}\;.
\end{split} \label{eq:diff_cond2}
\end{equation}
\end{widetext}
In Eq.~(\ref{eq:diff_cond2}) we applied the identity $g^R_{N,e} - g^{A}_{N,e} = 2{\rm i}\;{\rm Im}\left(g^R_{N,e}\right)$.
For simplicity let's suppose we have only one Andreev bound state (ABS) formed in the 
superconductor -- molecular core -- superconductor (S-mol-S) junction described by the Green's function $g^R_{\textrm{mol}}$.
In the presence of the normal lead, the ABS's starts to leak out via the normal lead resulting in the broadening of the ABS energy levels.
Since our main interest are the transport properties close to the mid of the HOMO-LUMO gap, in the relevant energy regime we do not expect any further bound states in $G^R_{\textrm{mol,mol}}$ besides the ones corresponding to the ABS's.
Thus, we might approximate $G^R_{\textrm{mol,mol}}$ as:
\begin{equation}
 G^R_{\textrm{mol,mol}}(E) \approx \frac{|ABS\rangle\langle ABS|}{E-E_{ABS} + {\rm i}\Gamma_{ABS}} \label{eq:ABS_GR}.
\end{equation}
Here the state $|ABS\rangle$ represents the wave function of the ABS in the molecule of energy $E_{ABS}$, and 
$\Gamma_{ABS}=\left\langle ABS\left| W_N^{\dagger}{\rm Im}\left(g^R_N\right)W_N \right|ABS\right\rangle$ is the level broadening originating from the escape rate of the particles through the normal lead.\cite{Claughton_1995}
The mathematical expression for $\Gamma_{ABS}$ calculates the overlap between the ABS wave function and the self energy of the normal lead. 
Thus, $\Gamma_{ABS}$ can be divided into two distinct terms, one related to the escape rate of the electron-like and the second one 
to the escape rate of the hole-like particles.
Namely, $\Gamma_{ABS} = \Gamma_{ABS,e} + \Gamma_{ABS,h}$, where:
\begin{equation}
 \Gamma_{ABS,e} = \left\langle ABS\left| W_N^{\dagger}\begin{pmatrix}
  {\rm Im}\left( g^{R}_{N,e}(E_{ABS})\right) & 0 \\ 0 & 0
 \end{pmatrix}W_N \right|ABS\right\rangle\;, \label{eq:Gamma_ABS_e}
\end{equation}
and
\begin{equation}
 \Gamma_{ABS,h} = \left\langle ABS\left| W_N^{\dagger}\begin{pmatrix}
  0 & 0 \\ 0 & {\rm Im}\left( g^{R}_{N,h}(E_{ABS})\right)
 \end{pmatrix}W_N \right|ABS\right\rangle\;. \label{eq:Gamma_ABS_h}
\end{equation}
Using the (\ref{eq:ABS_GR}) expression of $G^R_{\textrm{mol,mol}}$ and the invariance of the ${\rm Tr}(\dots)$ function against the cyclic permutation 
of its arguments one obtains for the differential conductance:
\begin{widetext}
\begin{equation} 
\begin{split}
   \frac{{\rm d} I^e_N}{{\rm d}V} \approx &-\frac{4e}{h}{\rm Im} \frac{\left\langle ABS\left| W_N^{\dagger}\begin{pmatrix}
  {\rm Im}\left( g^{R}_{N,e}\right) & 0 \\ 0 & 0
 \end{pmatrix}W_N \right|ABS\right\rangle}{eV-E_{ABS} + {\rm i}\Gamma_{ABS}} \\ 
   & - \frac{4e}{h}{\rm Im} \left\{ \frac{\left\langle ABS\left| W_N^{\dagger}\begin{pmatrix}
  {\rm Im}\left( g^{R}_{N,e}\right) & 0 \\ 0 & 0
 \end{pmatrix}W_N \right|ABS\right\rangle}{eV-E_{ABS} + {\rm i}\Gamma_{ABS}} 
 \frac{\left\langle ABS\left| W_N^{\dagger}g^A_N\tau_3W_N \right|ABS\right\rangle}{eV-E_{ABS} - {\rm i}\Gamma_{ABS}} \right\} \;.
\end{split} \label{eq:diff_cond3}
\end{equation}
\end{widetext}
Now making use of the definition of the broadening parameters $\Gamma_{ABS,e}$ and $\Gamma_{ABS,h}$ we end up with the following expression 
for the differential conductance:
\begin{equation}
 \frac{{\rm d} I^e_N}{{\rm d}V} \approx \frac{8e}{h}\frac{ \Gamma_{ABS,e}\Gamma_{ABS,h} }{(eV-E_{ABS})^2 + \Gamma_{ABS}^2}\;. 
 \label{eq:diff_cond_electron_final_simplified}
\end{equation}
In the above expression we neglected the energy dependence of the Green's function of the normal lead in a $\Gamma_{ABS}$ wide vicinity of the energy $E_{ABS}$. 
Accounting also for the hole-like part of the differential conductance (\ref{eq:diff_cond}) gives an additional factor of two in the final 
result due to the electron-hole symmetry. 
Thus, the total differential conductance would be given by Eq.~(7) of the main text.
In case we have more than one ABS in the junction, the first term of Eq.~(\ref{eq:diff_cond3}) would turn into a sum of Lorentzian resonances, 
while the second term evolves into a more complex mathematical expression: 
\begin{widetext}
\begin{equation}
 -\sum\limits_{p,q}{\rm Im} \left\{ \frac{\left\langle p\left| W_N^{\dagger}\begin{pmatrix}
  {\rm Im}\left( g^{R}_{N,e}\right) & 0 \\ 0 & 0
 \end{pmatrix}W_N \right|q\right\rangle}{eV-E_{p} + {\rm i}\Gamma_{pp}} 
 \frac{\left\langle q\left| W_N^{\dagger}g^A_N\tau_3W_N \right|p\right\rangle}{eV-E_{q} - {\rm i}\Gamma_{qq}} \right\} =
  {\rm Im}\sum\limits_{p,q} \frac{\Gamma_{pq,e}}{eV-E_{p} + {\rm i}\Gamma_{pp}} \frac{\Gamma_{qp,e}-\Gamma_{qp,h}}{eV-E_{q} - {\rm i}\Gamma_{qq}}\label{eq:crosstalk} 
\end{equation}
\end{widetext}
where $|q\rangle$, $E_q$ and $\Gamma_{q}$ represents the wave function, the energy and the broadening of the $q$th ABS, and the quantities $\Gamma_{qp,e}$ 
and $\Gamma_{qp,h}$ are defined similarly to Eqs.~(\ref{eq:Gamma_ABS_e}) and (\ref{eq:Gamma_ABS_h}), but the scalar product is taken between wave functions corresponding to different ABS's.
Besides regular Lorentzian resonances [$p=q$ terms of Eq.~(\ref{eq:crosstalk})] we see that the differential conductance is heavily influenced by the cross-talk of the individual ABS's.
Mathematically the product of two fractions on the right hand side of Eq.~(\ref{eq:crosstalk}) can be rewritten to a sum 
\begin{widetext}
\begin{equation}
 \frac{\Gamma_{pq,e}}{eV-E_{p} + {\rm i}\Gamma_{pp}} \frac{\Gamma_{qp,e}-\Gamma_{qp,h}}{eV-E_{q} - {\rm i}\Gamma_{qq}} = \frac{\lambda}{eV-E_{p} + {\rm i}\Gamma_{pp}} + \frac{\delta}{eV-E_{q} - {\rm i}\Gamma_{qq}}\;,
\end{equation}
\end{widetext}
where $\lambda$ and $\delta$ are in general complex numbers. 
(Individually both of them have singularity at $eV = (\Gamma_{pp}E_q+\Gamma_{qq}E_p)/(\Gamma_{pp}+\Gamma_{qq})$, but these singularities cancel 
each other in the sum of the two fractions.)
Consequently, the imaginary part of these fractions would differ from the regular Lorentzian function and the total differential conductance 
in the presence of multiple ABS's would be the sum of asymmetric Lorentzian resonances centered to the energies of the ABS's.
The asymmetry in the resonances is a signature of the cross-talk between the ABS's.

\section{Resonant oscillation}

As discussed in the main text, we try to explain the unconventional interference pattern by the interplay of the two paths depicted in Fig.~2 of the main text.
However the amplitude $t_{9,22}$ (defined by Eq.~(1) of the main text) might be expected to be much larger than the amplitude $t^{(9)}_{6,22}$ 
(defined by Eq.~(5) of the main text) which would suppress the interference effect between these two interfering paths.

The physical picture behind the small magnitude of $t^{(9)}_{6,22}$ relative to $t_{9,22}$ is associated to the particle transfer between the 
two superconducting banks.
The four tunneling processes between the molecular core and the superconducting electrodes significantly decreases the magnitude of 
the interfering amplitude $t^{(9)}_{6,22}$.
On the other hand, a resonant oscillation realized by the ABSs overwrites this physical picture. 
In this case the charge transport between the superconducting banks becomes resonantly amplified via the ABS and thus the amplitudes 
$t^{(9)}_{6,22}$ and $t_{9,22}$ becomes comparable. 
In summary, for energies close enough to the energy of an ABS the differential conductance shows an interference effect due to the resonant amplification 
of the interfering amplitude $t^{(9)}_{6,22}$, while for other energies the interference would be suppressed.

\section{Density of states}

In this subsection we give the technical details to calculate the density of states of the three-terminal molecular junction, which can be used to 
physically interpret the numerical results obtained by Eqs.~(4) of the main text and by Eq.~(\ref{eq:diff_cond}).
We calculate the density of states$\rho$  from the equilibrium Green's function of the three-terminal molecular junction 
labeled by $G^R_{\textrm{mol,mol}}$ in the calculations above. 
To be precise, $G^R_{\textrm{mol,mol}}$ labels only that block of the whole Green's function which contains only the molecular degrees of freedom.
Then the density of states can be defined as:
\begin{equation}
 \rho(E) = -\frac{1}{\pi}{\rm Tr}\left[{\rm Im}\left( G^R_{\textrm{mol,mol}}(E) \right)\right]\;.
\end{equation}
As for the differential conductance, $G^R_{\textrm{mol,mol}}$ can be calculated via the Dyson's equation (\ref{eq:Dyson_eq}) which is 
evaluated using the E\"otv\"os Quantum Utilities (EQuUs)\cite{EQUUS} software package.

\section{The tight-binding model of the molecular junctions}
\label{sec:tightbinding}

To describe the electrical transport processes in the studied molecular junctions we use a nearest neighbor tight binding model catching the dynamics of the $p_z$ electrons of the molecular core. 
\begin{figure}[h]
\begin{centering}
\includegraphics[trim={0 0 0 0},clip, width=0.9\linewidth]{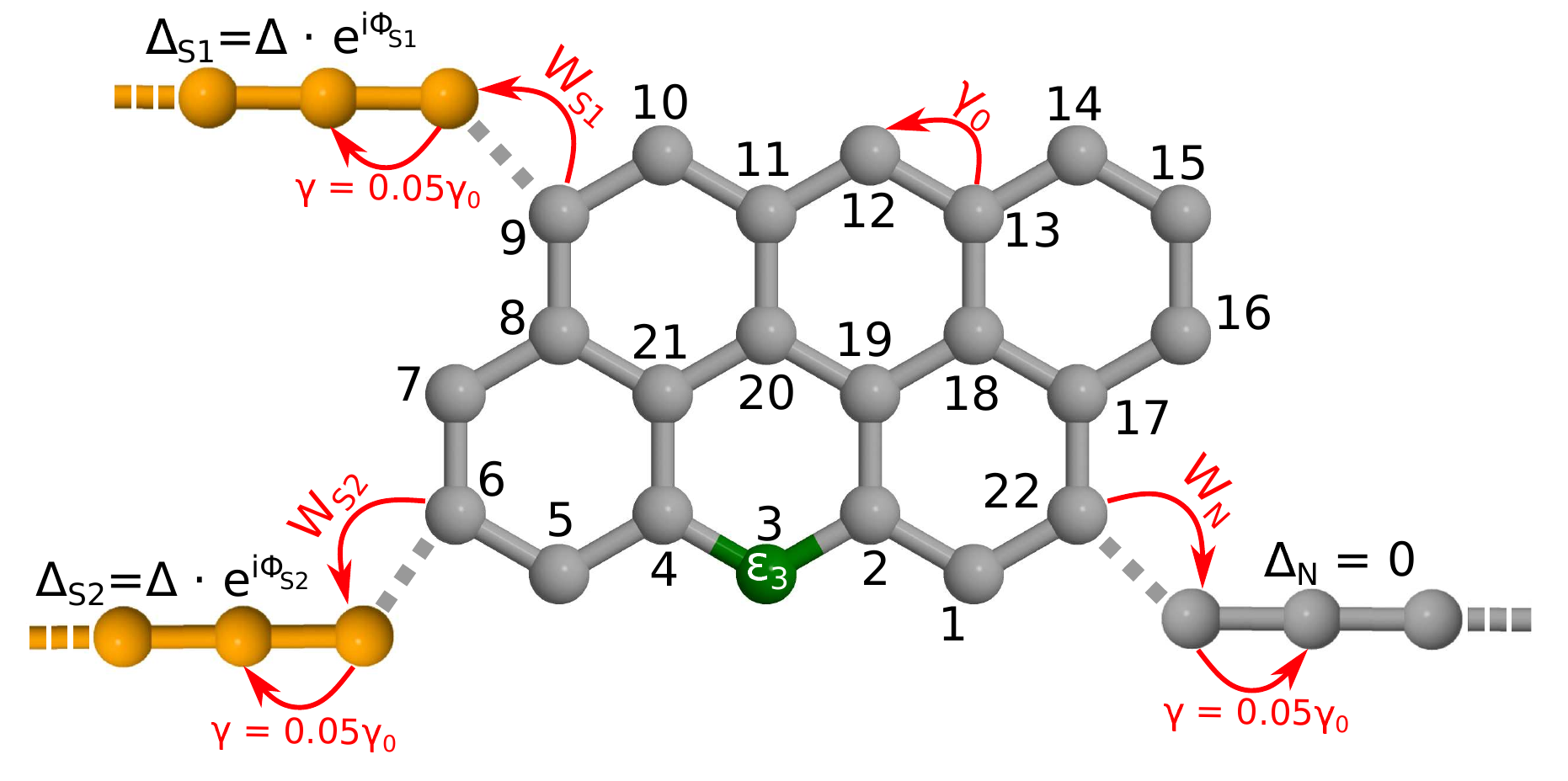}
\caption{The tight binding model of the Anthanthrene molecule attached to two superconductive and one normal lead. The sites in the molecular core are labeled by primed and unprimed numbers, while the hopping amplitude between the sites are characterized by a single number.
The normal and superconducting contacts are modeled by a one-dimensional conductive channels. }
\label{fig:TB_model}
\end{centering}
\end{figure}
The tight binding parameters describing the molecular core are chosen following the philosophy in Refs.\cite{Geng2015,Sangtarash2015}, where the aim is to highlight the role of connectivity in determining the transport properties of these molecular cores. For this reason,
the hopping integrals $\gamma_{ii'}=\gamma_0$ are set to unity and the on-site energies $\epsilon_i$ are set to zero. With other words, the unit
of energy is the hopping integral and the site energy is the energy origin. This means that the Hamiltonian of the molecule is simply a connectivity matrix and therefore all predicted effects are a result of connectivity alone.
The normal and superconducting contacts are modeled by a one-dimensional tight-binding chain. 
The transport properties of the junction have a weak dependence on the actual physical parameters of the leads as far as the leads remains metallic in the studied energy regime. 
Thus, we chose the physical parameters of the leads to increase the density of states in the leads and have the bandwidth of the conductive larger than the studied energy regime. 
In particular, we set the hopping amplitude in the contacts to $0.05\gamma_0$ and the on-site energy parameter to $0$. 
The superconducting contacts are modeled by an s-type superconducting pair potential $\Delta=0.001\gamma_0$.
(The pairing potential is zero anywhere else in the system.)
In the particular case the tight binding model of the Anthanthrene molecule connected to the superconducting and normal electrodes is shown in Fig.~\ref{fig:TB_model}.
Remarkably, as demonstrated in Refs.\cite{Geng2015,Sangtarash2015}, this approach yields the experimentally-measured conductance ratios of a range of PAHs.

Finally, as we explained in the main text, we tuned the transport properties of the molecular core by an inserting a substitutional heteroatom into the molecular core.
According to Ref.~\cite{Sangtarash2016}, the presence of the heteroatom have a strong influence on the inner quantum interference effects in the molecular core, even new conductive channels may open up in the molecular core.
In our theoretical model we account for the presence of a substitutional heteroatom by a modified 
on-site energy on a specific site in the molecule.

\section{Comparison of the local density of states on two molecular sites}
\label{sec:LDOS-on-two}

{As shown in  Fig.8(c) and (d) of the  the main text, which is reproduced below in Fig.~\ref{fig:site22}, the local density of states (LDOS) is suppressed for electron-like 
quasiparticles and enhanced for hole-like quasiparticles on molecular site $22$ (for the numbering of the molecular sites, see Fig.~\ref{fig:TB_model}). 
We have calculated the LDOS for the other sites of the molecular core as well and found that due to QI the LDOS of the electron and hole quasiparticles is different
on each site.  In particular, it can happen that, in contrast to  Fig.~\ref{fig:site22},  the  electron LDOS is larger than the hole LDOS. An example shown in 
Fig.~\ref{fig:site8}, where this asymmetry of LDOS can be clearly seen. }
\begin{figure}[h]
\begin{centering}
\includegraphics[trim={0 0 0 0},clip, width=1.1\linewidth]{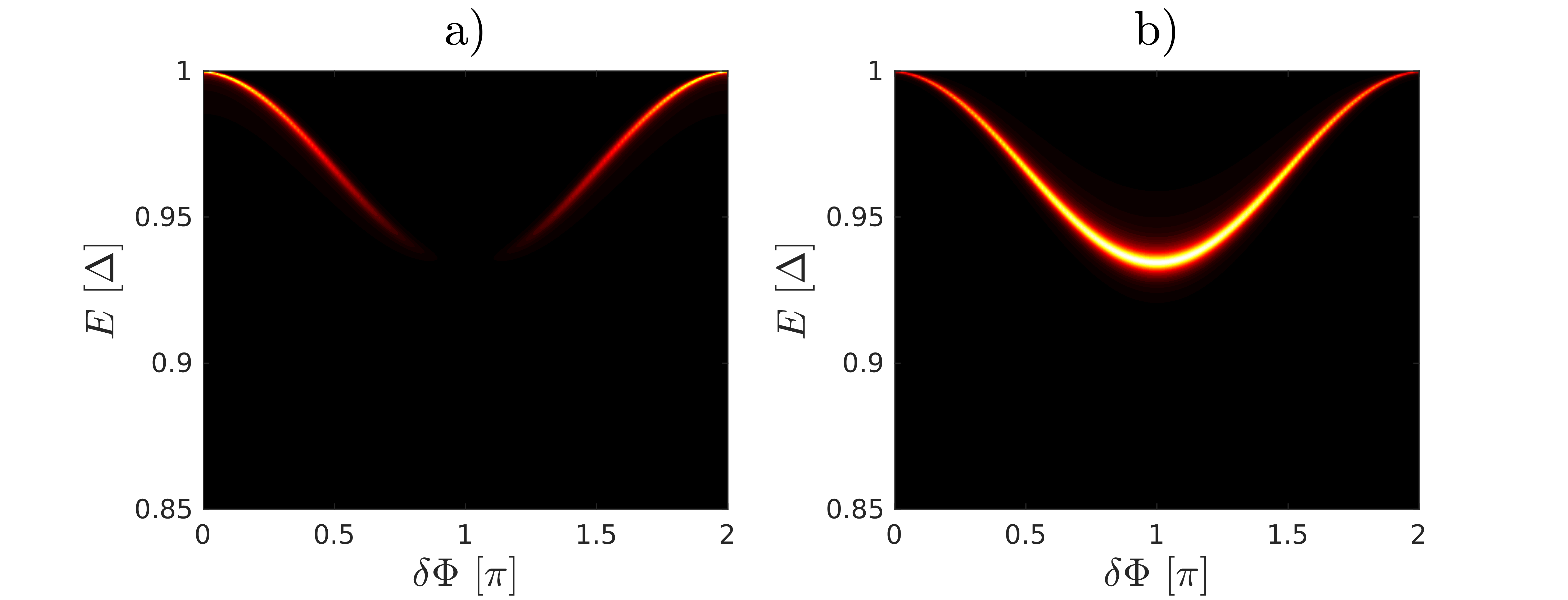}
\caption{The LDOS for electron (a) and hole (b) quasiparticles as a function of $\delta\Phi$ on molecular site  $22$  of the Andreev interferometer shown in 
Fig.6(c) of the main text and in Fig.~\ref{fig:TB_model}. In these calculations  $\epsilon_3=-0.50\gamma_0$. }
\label{fig:site22}
\end{centering}
\end{figure}

\begin{figure}[h]
\begin{centering}
\includegraphics[trim={0 0 0 0},clip, width=1.1\linewidth]{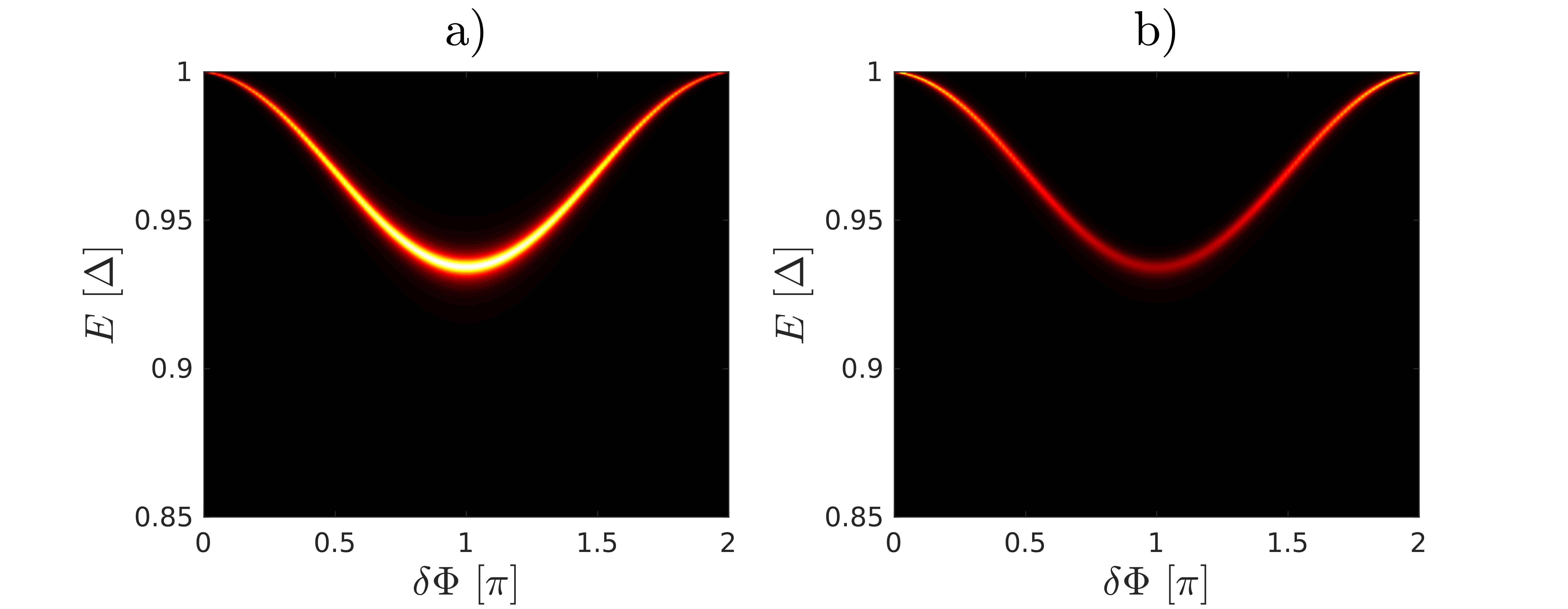}
\caption{The LDOS for electron (a) and hole (b) quasiparticles as a function of $\delta\Phi$ on molecular site  $8$ of the Andreev interferometer shown in 
Fig.6(c) of the main text and in Fig.~\ref{fig:TB_model}. In these calculations we used $\epsilon_3=-0.50\gamma_0$.  }
\label{fig:site8}
\end{centering}
\end{figure}
{As mentioned in the ``Conclusions and Outlook'' section of the main text, by attaching normal leads $N1$ and $N2$ to molecular sites  $8$ and $22$ and may enhance 
the non-local Andreev reflection $N1\rightarrow N2$ with respect to the local Andreev reflection $N1\rightarrow N1$.}

\bibliography{noneq-andreev-molec} 
\bibliographystyle{rsc} 

\end{document}